\documentclass[aps,a4paper,superscriptaddress,showpacs,preprintnumbers,amsmath,amssymb]{revtex4}

\usepackage{ulem}

\usepackage{psfrag} \usepackage{graphicx} \usepackage{dcolumn}
\usepackage{color} \usepackage{latexsym,amsfonts} \usepackage{bm}
\usepackage{amssymb}
\begin{document}

\title{On the correlation coefficient $T(E_e)$ \\ of the neutron beta
  decay, caused by the correlation structure\\ invariant under
  discrete P, C and T symmetries}

\author{A. N. Ivanov}\email{ivanov@kph.tuwien.ac.at}
\affiliation{Atominstitut, Technische Universit\"at Wien, Stadionallee
  2, A-1020 Wien, Austria}
\author{R. H\"ollwieser}\email{roman.hoellwieser@gmail.com}
\affiliation{Atominstitut, Technische Universit\"at Wien, Stadionallee
  2, A-1020 Wien, Austria}\affiliation{Department of Physics,
  Bergische Universit\"at Wuppertal, Gaussstr. 20, D-42119 Wuppertal,
  Germany} \author{N. I. Troitskaya}\email{natroitskaya@yandex.ru}
\affiliation{Atominstitut, Technische Universit\"at Wien, Stadionallee
  2, A-1020 Wien, Austria}
\author{M. Wellenzohn}\email{max.wellenzohn@gmail.com}
\affiliation{Atominstitut, Technische Universit\"at Wien, Stadionallee
  2, A-1020 Wien, Austria} \affiliation{FH Campus Wien, University of
  Applied Sciences, Favoritenstra\ss e 226, 1100 Wien, Austria}
\author{Ya. A. Berdnikov}\email{berdnikov@spbstu.ru}\affiliation{Peter
  the Great St. Petersburg Polytechnic University, Polytechnicheskaya
  29, 195251, Russian Federation}

\date{\today}

\begin{abstract}
  We analyze the correlation coefficient $T(E_e)$, which was
  introduced by Ebel and Feldman (Nucl. Phys. {\bf 4}, 213
  (1957)). The correlation coefficient $T(E_e)$ is induced by the
  correlation structure $(\vec{\xi}_n\cdot
  \vec{k}_{\bar{\nu}})(\vec{k}_e\cdot \vec{\xi}_e)/E_e E_{\bar{\nu}}$,
  where $\vec{\xi}_{n,e}$ are unit spin-polarization vectors of the
  neutron and electron, and $(E_{e,\bar{\nu}}, \vec{k}_{e,\bar{\nu}})$
  are energies and 3--momenta of the electron and antineutrino. Such a
  correlation structure is invariant under discrete P, C and T
  symmetries.  The correlation coefficient $T(E_e)$, calculated to
  leading order in the large nucleon mass $m_N$ expansion, is equal to
  $T(E_e) = - 2 g_A(1 + g_A)/(1 + 3 g^2_A) = - B_0$, i.e. of order
  $|T(E_e)| \sim 1$, where $g_A$ is the axial coupling
  constant. Within the Standard Model (SM) we describe the correlation
  coefficient $T(E_e)$ at the level of $10^{-3}$ by taking into the
  radiative corrections of order $O(\alpha/\pi)$ or the {\it outer}
  model-independent radiative corrections, where $\alpha$ is the
  fine-structure constant, and the corrections of order $O(E_e/m_N)$,
  caused by weak magnetism and proton recoil. We calculate also the
  contributions of interactions beyond the SM, including the
  contributions of the {\it second class} currents.
\end{abstract}

\maketitle

\section{Introduction}
\label{sec:introduction}

After the discovery by Chadwick in 1932 \cite{Chadwick1932}, neutron
has started to play an unprecedentedly important role in particle and
nuclear physics, astrophysics and cosmology
\cite{Abele2008}--\cite{Serebrov2016}. The experimental analysis of
correlation coefficients of the neutron beta decay allows to obtain
precise information about the structure of the Standard Model (SM)
interactions \cite{Lee1956, Lee1957a, Lee1957b, Severijns2011,
  DGH2014, PDG2020} and interactions beyond the SM \cite{Abele2008,
  Nico2009, Abele2016, Serebrov2017, Bodek2019} at low and high
energies \cite{Bhattacharya2012, Gardner2012, Cirigliano2012,
  Cirigliano2013a, Cirigliano2013b, Severijns2015, Bodek2016}.  For
the first time the most general form of the electron--energy and
angular distribution of the neutron beta decay for a polarized
neutron, a polarized electron and an unpolarized proton was proposed
by Jackson {\it et al.}  \cite{Jackson1957a, Jackson1957b,
  Jackson1958}. Following Jackson {\it et al.}  \cite{Jackson1957a}
the electron--energy and angular distribution of the neutron beta
decay for a polarized neutron, a polarized electron and an unpolarized
proton can be represented as follows
\begin{eqnarray}\label{eq:1}
\hspace{-0.15in}&&\frac{d^5 \lambda_n(E_e, \vec{k}_e,
  \vec{k}_{\bar{\nu}}, \vec{\xi}_n, \vec{\xi}_e)}{dE_e d\Omega_e
  d\Omega_{\bar{\nu}}} = (1 + 3 g^2_A)\,\frac{|G_V|^2}{16 \pi^5}\,(E_0
- E_e)^2 \,\sqrt{E^2_e - m^2_e}\, E_e\,F(E_e, Z =
1)\,\zeta(E_e)\,\Big\{1 + b(E_e)\,\frac{m_e}{E_e}\nonumber\\
\hspace{-0.15in}&& + a(E_e)\,\frac{\vec{k}_e\cdot
  \vec{k}_{\bar{\nu}}}{E_e E_{\bar{\nu}}} +
A(E_e)\,\frac{\vec{\xi}_n\cdot \vec{k}_e}{E_e} + B(E_e)\,
\frac{\vec{\xi}_n\cdot \vec{k}_{\bar{\nu}}}{E_{\bar{\nu}}} +
K_n(E_e)\,\frac{(\vec{\xi}_n\cdot \vec{k}_e)(\vec{k}_e\cdot
  \vec{k}_{\bar{\nu}})}{E^2_e E_{\bar{\nu}}}+
Q_n(E_e)\,\frac{(\vec{\xi}_n\cdot \vec{k}_{\bar{\nu}})(\vec{k}_e\cdot
  \vec{k}_{\bar{\nu}})}{ E_e E^2_{\bar{\nu}}}\nonumber\\
\hspace{-0.15in}&& + D(E_e)\,\frac{\vec{\xi}_n\cdot (\vec{k}_e\times
  \vec{k}_{\bar{\nu}})}{E_e E_{\bar{\nu}}} + G(E_e)\,\frac{\vec{\xi}_e
  \cdot \vec{k}_e}{E_e} + H(E_e)\,\frac{\vec{\xi}_e \cdot
  \vec{k}_{\bar{\nu}}}{E_{\bar{\nu}}} + N(E_e)\,\vec{\xi}_n\cdot
\vec{\xi}_e + Q_e(E_e)\,\frac{(\vec{\xi}_n\cdot \vec{k}_e)(
  \vec{k}_e\cdot \vec{\xi}_e)}{(E_e + m_e) E_e}\nonumber\\
\hspace{-0.15in}&& + K_e(E_e)\,\frac{(\vec{\xi}_e\cdot \vec{k}_e)(
  \vec{k}_e\cdot \vec{k}_{\bar{\nu}})}{(E_e + m_e)E_e E_{\bar{\nu}}} +
R(E_e)\,\frac{\vec{\xi}_n\cdot(\vec{k}_e \times \vec{\xi}_e)}{E_e} +
L(E_e)\,\frac{\vec{\xi}_e\cdot(\vec{k}_e \times
  \vec{k}_{\bar{\nu}})}{E_eE_{\bar{\nu}}}\Big\},
\end{eqnarray}
where we have used the notations in Refs.  \cite{Ivanov2013} --
\cite{Ivanov2020b}. Then, $g_A$ and $G_V$ are the axial and vector
coupling constants, respectively, \cite{PDG2020, Abele2008, Nico2009,
  Abele2018, Sirlin2018}, $\vec{\xi}_n$ and $\vec{\xi}_e$ are unit
spin--polarization vectors of the neutron and electron
\cite{Ivanov2013, Ivanov2017, Ivanov2019a} (see also
\cite{Itzykson1980}), respectively, $d\Omega_e$ and
$d\Omega_{\bar{\nu}}$ are infinitesimal solid angles in the directions
of electron $\vec{k}_e$ and antineutrino $\vec{k}_{\bar{\nu}}$
3--momenta, respectively, $E_0 = (m^2_n - m^2_p + m^2_e)/2m_n =
1.2926\,{\rm MeV}$ is the end--point energy of the electron--energy
spectrum \cite{Abele2008, Nico2009}, $F(E_e, Z = 1)$ is the
relativistic Fermi function, describing the electron--proton
final--state Coulomb interaction, is equal to (see, for example,
\cite{Blatt1952, Wilkinson1982} and a discussion in \cite{Ivanov2017})
\begin{eqnarray}\label{eq:2}
\hspace{-0.3in}F(E_e, Z = 1 ) = \Big(1 +
\frac{1}{2}\gamma\Big)\,\frac{4(2 r_pm_e\beta)^{2\gamma}}{\Gamma^2(3 +
  2\gamma)}\,\frac{\displaystyle e^{\,\pi \alpha/\beta}}{(1 -
  \beta^2)^{\gamma}}\,\Big|\Gamma\Big(1 + \gamma + i\,\frac{\alpha
}{\beta}\Big)\Big|^2,
\end{eqnarray}
where $\beta = k_e/E_e = \sqrt{E^2_e - m^2_e}/E_e$ is the electron
velocity, $\gamma = \sqrt{1 - \alpha^2} - 1$, $r_p$ is the electric
radius of the proton \cite{Antognini2013}. The correlation coefficient
$b(E_e)$ is the Fierz interference term \cite{Fierz1937}. The
structure and the value of the Fierz interference term may depend on
interactions beyond the SM \cite{Fierz1937}. An information of a
contemporary theoretical and experimental status of the Fierz
interference term can be found in \cite{ Hardy2020, Severijns2019,
  Abele2019, Young2019, Sun2020} (see also \cite{Ivanov2019y}). The
correlation coefficients of the electron-antineutrino correlations
$a(E_e)$, the electron asymmetry $A(E_e)$, the antineutrino asymmetry
$B(E_e)$ and others $G(E_e)$, $H(E_e)$, $Q_e(E_e)$ and $K_e(E_e)$
survive to leading order in the large nucleon mass $m_N$ expansion
\cite{Abele2008, Nico2009, Ivanov2017, Ivanov2019a} and depend on the
axial coupling constant $g_A$ only. The correlation coefficients
$a(E_e)$ and $Q_e(E_e)$ do not violate invariance under discrete
symmetries, i.e. parity conservation (P-invariance), time--reversal
invariance (T-invariance) and charge conjugation invariance
(C-invariance) \cite{Lee1957b}. This is unlike the correlation
coefficients $A(E_e)$, $B(E_e)$, $G(E_e)$, $H(E_e)$ and $K_e(E_e)$,
which provide a quantitative information about effects of parity
violation (P-odd effects) \cite{Jackson1957b}. In turn, the
correlation coefficients $K_n(E_e)$ and $Q_n(E_e)$ appear only to
next-to-leading order in the large nucleon mass $m_N$ expansion.  They
are caused by the contributions of weak magnetism and proton recoil
\cite{Bilenky1959} (see also \cite{Wilkinson1982, Ando2004,
  Gudkov2006} and \cite{Ivanov2013}) and measure the strength of P-odd
effects.  The correlation coefficients $D(E_e)$, $R(E_e)$ and $L(E_e)$
characterize quantitatively the strength of effects of violation of
time-reversal invariance (T-odd effects). In addition, the correlation
coefficients $D(E_e)$ and $L(E_e)$ are responsible also for violation
of charge invariance (C--odd effects), whereas the correlation
coefficient $R(E_e)$ defines also quantitatively effects of violation
of parity invariance (P-odd effects) \cite{Jackson1957a,
  Jackson1957b}. In the SM the correlation coefficients $D(E_e)$,
$R(E_e)$ and $L(E_e)$ are induced by the distortion of the Dirac wave
function of the electron in the Coulomb field of the proton
\cite{Jackson1957b, Jackson1958, Callan1967, Ando2009} (see also
\cite{Ivanov2017, Ivanov2019a}). The radiative corrections of order
$O(\alpha/\pi)$, where $\alpha$ is the fine--structure constant
\cite{PDG2020}, to the neutron lifetime and correlation coefficients
in Eq.(\ref{eq:1}) were calculated in \cite{Sirlin1967} --
\cite{Hayen2020} (see also \cite{Ando2004, Gudkov2006} and
\cite{Ivanov2013, Ivanov2017, Ivanov2019a}). The corrections of order
$O(E_e/m_N)$, caused by weak magnetism and proton recoil, were
calculated in \cite{Bilenky1959}, \cite{Wilkinson1982, Ando2004,
  Gudkov2006} (see also \cite{Ivanov2013, Ivanov2017,
  Ivanov2019a}). So one may argue that within the SM the correlation
coefficients in Eq.(\ref{eq:1}) were fully investigated theoretically
to order $10^{-3}$, caused by radiative corrections of order
$O(\alpha/\pi)$ and the corrections of order $O(E_e/m_N)$, induced by
weak magnetism and proton recoil.

The electron-energy and angular distribution in Eq.(\ref{eq:1}) was
supplemented by the term with the correlation structure
$(\vec{\xi}_n\cdot \vec{k}_{\bar{\nu}})(\vec{k}_e\cdot
\vec{\xi}_e)/E_e E_{\bar{\nu}}$ and the correlations coefficient
$T(E_e)$ by Ebel and Feldman \cite{Ebel1957}. Such a correlation
structure is invariant under discrete P, C and T symmetries
\cite{Lee1957b, Jackson1957a, Jackson1957b}. The main peculiarity of
the correlation coefficient $T(E_e)$ in comparison to other
correlation coefficients, introduced by Ebel and Feldman
\cite{Ebel1957} in addition to the correlation coefficients proposed
by Jackson {\it et al.}  \cite{Jackson1957a} in Eq.(\ref{eq:1}), is to
be finite with the absolute value of order of 1, i.e. $|T(E_e)| \sim
1$, to leading order in the large nucleon mass $m_N$ expansion.  This
paper is addressed to the detailed analysis of the structure of the
correlation coefficient $T(E_e)$.

The paper is organized as follows. In section \ref{sec:LO} within the
standard effective $V - A$ theory of weak interactions
\cite{Feynman1958} we calculate the correlation coefficient $T(E_e)$
to leading order in the large nucleon mass $m_N$ expansion. Such a
correlation coefficient was introduced by Ebel and Feldman
\cite{Ebel1957}, who calculated the correlation coefficient by using
the effective phenomenological interactions by Jackson {\it et al.}
\cite{Jackson1957a, Jackson1957b} only.  We show that the absolute
value of $T(E_e)$ is of order of 1, i.e. $|T(E_e)| \sim 1$. According
to \cite{Jackson1957a, Jackson1957b}, the correlation structure
responsible for $T(E_e)$ does not violate invariance under discrete P,
C and T symmetries \cite{Lee1957b}. In section \ref{sec:NLO} within
the SM we give a detailed description of the correlation coefficient
$T(E_e)$ at the level of $10^{-3}$ including the radiative corrections
of order $O(\alpha/\pi)$, calculated to leading order in the large
nucleon mass $m_N$ expansion, and the corrections of order
$O(E_e/m_N)$, caused by weak magnetism and proton recoil. In section
\ref{sec:BSM} we calculate the contribution of interactions beyond the
SM expressed in terms of the phenomenological coupling constants of
the effective low-energy weak interactions proposed by Jackson {\it et
  al.}  \cite{Jackson1957a}. In section \ref{sec:Gparity} we calculate
the contribution of the {\it second class} currents or the $G$--odd
correlations (regarding $G$--parity invariance of strong interactions,
we refer to the paper by Lee and Yang \cite{Lee1956a}) by Weinberg
\cite{Weinberg1958}. In section \ref{sec:Abschluss} we discuss the
obtained results and perspectives of i) an experimental analysis of
the correlation coefficient $T(E_e)$ and ii) an improvement of the SM
description of $T(E_e)$ at the level of $10^{-5}$. In section
\ref{sec:appendix} we give a detailed calculation of the contribution
of the neutron radiative beta decay $n \to p + e^- + \bar{\nu}_e +
\gamma$ to the correlation coefficient $T(E_e)$ providing a removal of
the dependence of the radiative corrections on the infrared cut--off.

\section{The structure of the correlation coefficient $T(E_e)$  to
  leading order in the large nucleon mass expansion in the effective
  $V - A$ theory}
\label{sec:LO}

The analysis of the electron--energy and angular distribution of the
neutron beta decay for a polarized neutron, a polarized electron and
an unpolarized proton within the standard effective $V - A$ theory of
weak low-energy interactions, carried out even to leading order in the
large nucleon mass expansion, shows that the set of correlation
coefficients in Eq.(\ref{eq:1}), which survive in such a limit, is not
complete. In order to show this we use the standard effective
Lagrangian of $V - A$ low--energy weak interactions \cite{Feynman1958}
\begin{eqnarray}\label{eq:3}
\hspace{-0.3in}{\cal L}_W(x) = - G_V\,[\bar{\psi}_p(x)\gamma_{\mu}(1 -
  g_A \gamma^5)\psi_n(x)][\bar{\psi}_e(x)\gamma^{\mu} (1 - \gamma^5)
  \psi_{\nu}(x)],
\end{eqnarray}
 where $G_V$ and $g_A$ are the vector and axial coupling constants
 \cite{PDG2020, Abele2008, Nico2009}, $\psi_p(x)$, $\psi_n(x)$,
 $\psi_e(x)$ and $\psi_{\nu}(x)$ are the field operators of the
 proton, neutron, electron and antineutrino, respectively,
 $\gamma^{\mu} = (\gamma^0, \vec{\gamma}\,)$ and $\gamma^5$ are the
 Dirac matrices \cite{Itzykson1980}. In the non-relativistic
 approximation for the neutron and proton the amplitude of the neutron
 beta decay, calculated with the effective Lagrangian Eq.(\ref{eq:3}),
 is \cite{Ivanov2013}
 \begin{eqnarray}\label{eq:4}
\hspace{-0.3in}M(n \to p e^- \bar{\nu}_e) = - 2 m_n G_V
\Big\{[\varphi^{\dagger}_p\varphi_n]\,[\bar{u}_e \gamma^0 (1 -
  \gamma^5) v_{\bar{\nu}}] + g_A [\varphi^{\dagger}_p \vec{\sigma}\,
  \varphi_n] \cdot [\bar{u}_e \vec{\gamma}\, (1 - \gamma^5)
  v_{\bar{\nu}}]\Big\},
\end{eqnarray}
where $\varphi_n$ and $\varphi_p$ are Pauli wave functions of the
neutron and proton, whereas $u_e$ and $v_{\bar{\nu}}$ are Dirac wave
functions of the electron and antineutrino, $\vec{\sigma}$ are Pauli
$2 \times 2$ matrices of the neutron spin
\cite{Itzykson1980}. According to \cite{Ivanov2013}, the
electron--energy and angular distribution of the neutron beta decay
for a polarized neutron, a polarized electron and an unpolarized
proton, calculated with the effective Lagrangian Eq.(\ref{eq:3}), is
given by
 \begin{eqnarray}\label{eq:5}
\hspace{-0.3in}&&\frac{d^5\lambda_n(E_e, \vec{k}_e,
  \vec{k}_{\bar{\nu}},\vec{\xi}_n, \vec{\xi}_e)}{ d E_e
  d\Omega_ed\Omega_{\bar{\nu}}} = (1 + 3 g^2_A) \, \frac{|G_V|^2}{16
  \pi^5}\,(E_0 - E_e)^2\,\sqrt{E^2_e - m^2_e}\,E_e\,F(E_e, Z = 1)
\sum_{\rm pol.}\frac{|M(n \to p e^- \bar{\nu}_e)|^2}{(1 + 3 g^2_A)
  |G_V|^2 64 m^2_n E_e E_{\bar{\nu}}},\nonumber\\
\hspace{-0.3in}&&
\end{eqnarray}
where we sum over polarizations of the massive fermions. The sum over
polarizations of the massive fermions is defined by \cite{Ivanov2013,
  Ivanov2017, Ivanov2019a}
\begin{eqnarray}\label{eq:6}
\hspace{-0.3in}&&\sum_{\rm pol.}\frac{|M(n \to p e^-
  \bar{\nu}_e)|^2}{(1 + 3 g^2_A)|G_V|^2 64 m^2_n E_e E_{\bar{\nu}}} =
\frac{1}{(1 + 3 g^2_A) 8 E_e E_{\bar{\nu}}}\Big\{{\rm tr}\{(1 +
\vec{\xi}_n\cdot \vec{\sigma}\,)\}\,{\rm tr}\{(\hat{k}_e + m_e
\gamma^5 \hat{\zeta}_e) \gamma^0 \hat{k}_{\bar{\nu}}\gamma^0 (1 -
\gamma^5)\} \nonumber\\
\hspace{-0.3in}&& + g_A {\rm tr}\{(1 + \vec{\xi}_n\cdot
\vec{\sigma}\,) \vec{\sigma}\,\}\cdot {\rm tr}\{(\hat{k}_e + m_e
\gamma^5 \hat{\zeta}_e) \gamma^0 \hat{k}_{\bar{\nu}}\vec{\gamma}\, (1
- \gamma^5)\} + g_A {\rm tr}\{(1 + \vec{\xi}_n\cdot \vec{\sigma}\,)
\vec{\sigma}\,\}\cdot {\rm tr}\{(\hat{k}_e + m_e \gamma^5
\hat{\zeta}_e) \vec{\gamma}\, \hat{k}_{\bar{\nu}}\gamma^0 (1 -
\gamma^5)\} \nonumber\\
\hspace{-0.3in}&& + g^2 _A{\rm tr}\{(1 + \vec{\xi}_n\cdot
\vec{\sigma}\,) \sigma^a\sigma^b \} {\rm tr}\{(\hat{k}_e + m_e
\gamma^5 \hat{\zeta}_e) \gamma^b \hat{k}_{\bar{\nu}}\gamma^a (1 -
\gamma^5)\}\Big\},
\end{eqnarray}
where $\zeta_e$ is the 4--vector of the spin--polarization of the
electron. It is defined by \cite{Itzykson1980}
\begin{eqnarray}\label{eq:7}
\zeta_e = (\zeta^0_e, \vec{\zeta}_e) = \Big(\frac{\vec{k}_e\cdot
  \vec{\xi}_e}{m_e}, \vec{\xi}_e + \frac{(\vec{k}_e \cdot \vec{\xi}_e)
  \vec{k}_e }{m_e(E_e + m_e)}\Big).
\end{eqnarray}
The 4-vector $\zeta_e$ of the spin--polarization of the electron is
normalized by $\zeta^2_e = - 1$. It obeys also the constraint
$k_e\cdot \zeta_e = 0$ \cite{Itzykson1980}. Calculating the traces
over the nucleon degrees of freedom and using the properties of the
Dirac matrices \cite{Itzykson1980}
\begin{eqnarray}\label{eq:8}
\hspace{-0.3in}\gamma^{\alpha}\gamma^{\nu}\gamma^{\mu} =
\gamma^{\alpha}\eta^{\nu\mu} - \gamma^{\nu}\eta^{\mu\alpha} +
\gamma^{\mu}\eta^{\alpha\nu} +
i\,\varepsilon^{\alpha\nu\mu\beta}\,\gamma_{\beta}\gamma^5,
\end{eqnarray}
where $\eta^{\mu\nu}$ is the metric tensor of the Minkowski
space--time, $\varepsilon^{\alpha\nu\mu\beta}$ is the Levi--Civita
tensor defined by $\varepsilon^{0123} = 1$ and
$\varepsilon_{\alpha\nu\mu\beta}= - \varepsilon^{\alpha\nu\mu\beta}$
\cite{Itzykson1980}, we transcribe the right-hand-side (r.h.s.) of
Eq.(\ref{eq:6}) into the form \cite{Ivanov2013, Ivanov2017,
  Ivanov2019a}
\begin{eqnarray}\label{eq:9}
\hspace{-0.3in}\sum_{\rm pol.}\frac{|M(n \to p e^- \bar{\nu}_e)|^2}{(1
  + 3 g^2_A)|G_V|^2 64 m^2_n E_e E_{\bar{\nu}}} &=& \frac{1}{ 4
  E_e}\Big\{\Big(1 + B_0\, \frac{\vec{\xi}_n\cdot
  \vec{k}_{\bar{\nu}}}{E_{\bar{\nu}}}\Big){\rm tr}\{(\hat{k}_e + m_e
\gamma^5 \hat{\zeta}_e) \gamma^0 (1 - \gamma^5)\} \nonumber\\
\hspace{-0.3in}&& + \Big(A_0\, \vec{\xi}_n +
a_0\,\frac{\vec{k}_{\bar{\nu}}}{E_{\bar{\nu}}}\Big) \cdot {\rm
  tr}\{(\hat{k}_e + m_e \gamma^5 \hat{\zeta}_e)
\vec{\gamma} (1 - \gamma^5)\}\Big\},
\end{eqnarray}
where $a_0$, $A_0$ and $B_0$ are defined in terms of the axial
coupling constant $g_A$ only \cite{Abele2008, Nico2009}
\begin{eqnarray}\label{eq:10}
\hspace{-0.3in} a_0 = \frac{1 - g^2_A}{1 + 3 g^2_A}\quad,\quad A_0 = 2
\frac{g_A(1 - g_A)}{1 + 3 g^2_A}\quad,\quad B_0 = 2 \frac{g_A(1 +
  g_A)}{1 + 3 g^2_A}.
\end{eqnarray}
Having calculated the traces over lepton degrees of freedom, we arrive
at the expression
\begin{eqnarray}\label{eq:11}
\hspace{-0.3in}&&\sum_{\rm pol.}\frac{|M(n \to p e^-
  \bar{\nu}_e)|^2}{(1 + 3 g^2_A)|G_V|^2 64 m^2_n E_e E_{\bar{\nu}}} =
\Big(1 + B_0\, \frac{\vec{\xi}_n\cdot
  \vec{k}_{\bar{\nu}}}{E_{\bar{\nu}}}\Big)\Big(1 -
\frac{\vec{\xi}_e\cdot \vec{k}_e}{E_e}\Big) + \Big(A_0\, \vec{\xi}_n +
a_0\,\frac{\vec{k}_{\bar{\nu}}}{E_{\bar{\nu}}}\Big)
\cdot\Big(\frac{\vec{k}_e}{E_e} - \frac{m_e}{E_e}\,\vec{\xi}_e
\nonumber\\
\hspace{-0.3in}&& - \frac{\vec{k}_e(\vec{k}_e \cdot \vec{\xi}_e)}{(E_e
  + m_e) E_e}\Big)= 1 + a_0\, \frac{\vec{k}_e \cdot
  \vec{k}_{\bar{\nu}}}{E_e E_{\bar{\nu}}} + A_0\, \frac{\vec{\xi}_n
  \cdot \vec{k}_e}{E_e} + B_0\, \frac{\vec{\xi}_n \cdot
  \vec{k}_{\bar{\nu}}}{E_{\bar{\nu}}} - \frac{\vec{\xi}_e \cdot
  \vec{k}_e}{E_e} - a_0\, \frac{m_e}{E_e}\, \frac{\vec{\xi}_e \cdot
  \vec{k}_{\bar{\nu}}}{E_{\bar{\nu}}} - A_0\, \frac{m_e}{E_e}\,
\vec{\xi}_n \cdot \vec{\xi}_e \nonumber\\
\hspace{-0.3in}&& - A_0\, \frac{(\vec{\xi}_n \cdot \vec{k}_e)(
  \vec{k}_e \cdot \vec{\xi}_e)}{(E_e + m_e) E_e} - a_0\,
\frac{(\vec{\xi}_e \cdot \vec{k}_e)( \vec{k}_e \cdot
  \vec{k}_{\bar{\nu}})}{(E_e + m_e) E_e E_{\bar{\nu}}} - B_0 \,
\frac{(\vec{\xi}_n \cdot \vec{k}_{\bar{\nu}})(\vec{\xi}_e \cdot
  \vec{k}_e)}{E_e E_{\bar{\nu}}}.
\end{eqnarray}
Plugging Eq.(\ref{eq:11}) into Eq.(\ref{eq:5}) we obtain the
electron--energy and angular distribution of the neutron beta decay,
calculated to leading order in the large nucleon mass $m_N$ expansion
. We get
\begin{eqnarray}\label{eq:12}
\hspace{-0.3in}&&\frac{d^5\lambda_n(E_e, \vec{k}_e,
  \vec{k}_{\bar{\nu}},\vec{\xi}_n, \vec{\xi}_e)}{ d E_e
  d\Omega_ed\Omega_{\bar{\nu}}} = (1 + 3 g^2_A) \, \frac{|G_V|^2}{16
  \pi^5}\,(E_0 - E_e)^2 \,\sqrt{E^2_e - m^2_e}\,E_e\,F(E_e, Z = 1)
\nonumber\\
\hspace{-0.3in}&& \times \Big\{1 + a_0\, \frac{\vec{k}_e \cdot
  \vec{k}_{\bar{\nu}}}{E_e E_{\bar{\nu}}} + A_0\, \frac{\vec{\xi}_n
  \cdot \vec{k}_e}{E_e} + B_0\, \frac{\vec{\xi}_n \cdot
  \vec{k}_{\bar{\nu}}}{E_{\bar{\nu}}} - \frac{\vec{\xi}_e \cdot
  \vec{k}_e}{E_e} - a_0\, \frac{m_e}{E_e}\, \frac{\vec{\xi}_n \cdot
  \vec{k}_{\bar{\nu}}}{E_{\bar{\nu}}} - A_0\, \frac{m_e}{E_e}\,
\vec{\xi}_n \cdot \vec{\xi}_e \nonumber\\
\hspace{-0.3in}&& - A_0\, \frac{(\vec{\xi}_n \cdot \vec{k}_e)(
  \vec{k}_e \cdot \vec{\xi}_e)}{(E_e + m_e) E_e} - a_0\,
\frac{(\vec{\xi}_e \cdot \vec{k}_e)( \vec{k}_e \cdot
  \vec{k}_{\bar{\nu}})}{(E_e + m_e) E_e E_{\bar{\nu}}} - B_0 \,
\frac{(\vec{\xi}_n \cdot \vec{k}_{\bar{\nu}})(\vec{\xi}_e \cdot
  \vec{k}_e)}{E_e E_{\bar{\nu}}}\Big\}.
\end{eqnarray}
From the comparison of Eq.(\ref{eq:12}) with Eq.(\ref{eq:1}) we
determine the following correlation coefficients in the
electron--energy and angular distribution, calculated to leading order
in the large nucleon mass $m_N$ expansion:
\begin{eqnarray}\label{eq:13}
\hspace{-0.3in}a(E_e) &=& a_0\quad,\quad A(E_e) = A_0 \quad,\quad
B(E_e) = B_0 \quad,\quad G(E_e) = - 1\,, \nonumber\\
\hspace{-0.3in}H(E_e) &=& - a_0\, \frac{m_e}{E_e}\;,\; N(E_e) =
- A_0\, \frac{m_e}{E_e}\;, \; Q_e(E_e) = - A_0 \;,\;
K_e(E_e) = - a_0.
\end{eqnarray}
The correlation structure of the last term in Eq.(\ref{eq:12}) was
introduced by Ebel and Feldman \cite{Ebel1957} with the correlation
coefficient $T(E_e)$.  According to our analysis carried out above to
leading order in the large nucleon mass expansion, it is equal to
$T(E_e) = - B_0$. The correlation structure responsible for the
correlation coefficient $T(E_e)$ is invariant under discrete P, C and
T symmetries \cite{Lee1957b, Jackson1957a, Jackson1957b,
  Itzykson1980}.

Having calculated the correlation coefficient $T(E_e)$ to leading
order in the large nucleon mass expansion and without radiative
corrections, we may proceed to the analysis of the structure of this
correlation coefficient by taking into account the radiative
corrections of order $O(\alpha/\pi)$ or so-called {\it outer}
model-independent radiative corrections \cite{Wilkinson1970} and
corrections of order $O(E_e/m_N)$, caused by weak magnetism and proton
recoil. This should give within the SM the theoretical description of
the correlation coefficient $T(E_e)$ at the level of $10^{-3}$.

\section{The correlation coefficient $T(E_e)$ to order $10^{-3}$}
\label{sec:NLO}

The procedure of the calculation of the radiative corrections of order
$O(\alpha/\pi)$ and the corrections of order $O(E_e/m_N)$, caused by
weak magnetism and proton recoil, has been expounded well in
\cite{Ivanov2013, Ivanov2017, Ivanov2019a}. For the calculation of
these corrections we use the following effective Lagrangian
\cite{Ivanov2013, Ivanov2017, Ivanov2019a}
\begin{eqnarray}\label{eq:14}
\hspace{-0.3in}{\cal L}_{\rm W\gamma}(x) &=& -
G_V\,\Big\{[\bar{\psi}_p(x)\gamma_{\mu}(1 - g_A \gamma^5)\psi_n(x)] +
\frac{\kappa}{2 m_N}
\partial^{\nu}[\bar{\psi}_p(x)\sigma_{\mu\nu}\psi_n(x)]\Big\}
        [\bar{\psi}_e(x)\gamma^{\mu}(1 -
          \gamma^5)\psi_{\nu}(x)]\nonumber\\
\hspace{-0.3in}&& - e\, \big( \bar{\psi}_p(x)\gamma^{\mu}\psi_p(x) -
\bar{\psi}_e(x)\gamma^{\mu}\psi_e(x) \big)\, A_{\mu}(x).
\end{eqnarray}
In comparison to the effective Lagrangian Eq.(\ref{eq:3}), we have
added i) in the hadronic current an additional term with the Lorentz
structure, defined the Dirac matrix $\sigma^{\mu\nu} =
\frac{i}{2}(\gamma^{\mu}\gamma^{\nu} - \gamma^{\nu}\gamma^{\mu})$
\cite{Itzykson1980}, where $\kappa = \kappa_p - \kappa_n = 3.7059$ is
the isovector anomalous magnetic moment of the nucleon, induced by the
anomalous magnetic moments of the proton $\kappa_p = 1.7929$ and the
neutron $\kappa_n = - 1.9130$ and measured in nuclear magneton
\cite{PDG2020}, and ii) the electromagnetic interactions for the
proton and electron, where $e$ is the electric charge of the proton
and $A_{\mu}(x)$ is the 4--vector potential of the electromagnetic
field.

Using the amplitude of the neutron beta decay (see Eq.(D-52) in
Appendix D of \cite{Ivanov2013} with a replacement $\tilde{\lambda} =
- \tilde{g}_A$, where $\tilde{g}_A = g_A(1 - E_0/2 m_N)$), following
\cite{Ivanov2013, Ivanov2019a} and skipping intermediate calculations
we obtain
\begin{eqnarray}\label{eq:15}
\hspace{-0.3in} &&\zeta(E_e) T(E_e) = - B_0\, \Big(1 +
\frac{\alpha}{\pi}\, f_{\beta^-_c}(E_e, \mu)\Big) + \frac{1}{1 + 3
  g^2_A}\,\frac{E_0}{m_N}\,\Big(2 g_A \big(g_A + (\kappa + 1)\big) -
\big(7g^2_A + g_A(3\kappa + 8) + (\kappa + 1)\big)\,
\frac{E_e}{E_0}\Big),\nonumber\\
\hspace{-0.3in}&&
\end{eqnarray}
where we have taken into account the contribution of the phase volume
of the neutron beta decay \cite{Ivanov2013, Ivanov2017, Ivanov2019a,
  Ivanov2020b}.  The function $f_{\beta^-_c}(E_e, \mu)$ was calculated
by Sirlin \cite{Sirlin1967} (for the calculation of the function
$f_{\beta^-_c}(E_e, \mu)$ in details we refer to \cite{Ivanov2013}),
$\mu$ is a covariant infrared cut-off having a meaning of the photon
mass \cite{Sirlin1967}. For the definition of the function
$\zeta(E_e)$ we refer to \cite{Jackson1957a, Jackson1957b}, where the
function $\zeta(E_e)$ was calculated to leading order in the large
nucleon mass $m_N$ expansion. However, below we shall use the function
$\zeta(E_e)$, calculated in \cite{Ivanov2013} to order $10^{-3}$ by
taking into account the contributions of the radiative corrections of
order $O(\alpha/\pi)$ and the corrections of order $O(E_e/m_N)$,
caused by weak magnetism and proton recoil (see also
\cite{Gudkov2006}). According to \cite{Wilkinson1970}, the function
$f_{\beta^-_c}(E_e, \mu)$ defines the contribution of
model--independent or {\it outer} radiative corrections. Following
\cite{Ivanov2013} the function $f_{\beta^-_c}(E_e, \mu)$ can be
rewritten as follows
\begin{eqnarray}\label{eq:16}
  \hspace{-0.3in} f_{\beta^-_c}(E_e, \mu) = g_n(E_e) + \frac{1 -
    \beta^2}{2\beta}\,{\ell n}\frac{1 + \beta}{1 - \beta} -
  g^{(1)}_{\beta^-_c\gamma}(E_e, \mu),
\end{eqnarray}
where $2 g_n(E_e)$ is Sirlin's function, defining the {\it outer}
radiative corrections of order $O(\alpha/\pi)$ to the neutron lifetime
\cite{Sirlin1967}. The function $g^{(1)}_{\beta^-_c\gamma}(E_e, \mu)$
corresponds to the contribution of the neutron radiative beta decay $n
\to p + e^- + \bar{\nu}_e + \gamma$ with a real photon $\gamma$, which
should be added, according to Berman \cite{Berman1958} and Kinoshita
and Sirlin \cite{Kinoshita1959} (see also Sirlin \cite{Sirlin1967}),
for the removal of the dependence of the neutron lifetime on the
infrared cut--off. Following \cite{Ivanov2013} the function
$g^{(1)}_{\beta^-_c\gamma}(E_e, \mu)$ is defined by the integral
\begin{eqnarray}\label{eq:17}
\hspace{-0.3in}g^{(1)}_{\beta^-_c\gamma}(E_e, \mu) =
\int\frac{d\Omega_{\gamma}}{4\pi} \int^{E_0 - E_e}_{\mu}\frac{dq_0
}{q_0}\, \sqrt{1 - \frac{\mu^2}{q^2_0}}\, \frac{(E_0 - E_e -
  q_0)^2}{(E_0 - E_e)^2}\,\Big\{\frac{k^2_e - (\vec{v}\cdot
  \vec{k}_e)^2}{(E_e - \vec{v}\cdot \vec{k}_e)^2}\Big(1 +
\frac{q_0}{E_e}\Big) + \frac{1}{E_e - \vec{v}\cdot
  \vec{k}_e}\,\frac{q^2_0}{E_e}\Big\},
\end{eqnarray}
where $q_0 = \sqrt{q^2 + \mu^2}$, $q = |\vec{q}\,|$ and $\vec{v} =
\vec{q}/q_0$. The analytical expression of the function
$g^{(1)}_{\beta^-_c\gamma}(E_e, \mu)$, which was calculated in
\cite{Ivanov2013}, we adduce in section \ref{sec:appendix} for
completeness. Then, in Eq.(\ref{eq:17}) we integrate over directions
of the 3--momentum $\vec{q}$ of the photon \cite{Ivanov2013}. As has
been shown in \cite{Ivanov2013, Ivanov2017, Ivanov2019a} for the
calculation of the radiative corrections to the correlation
coefficients the function $g^{(1)}_{\beta^-_c\gamma}(E_e, \mu)$ can be
also regularized by a non-covariant infrared cut-off $\omega_{\rm
  min}$. Setting $\mu = 0$ in Eq.(\ref{eq:17}) and integrating over
directions of the 3--momentum of the photon we arrive at the integral
\cite{Ivanov2013} (see Eq.(B-15) in Appendix B in
Ref. \cite{Ivanov2013})
\begin{eqnarray}\label{eq:18}
\hspace{-0.3in}g^{(1)}_{\beta^-_c\gamma}(E_e,\omega_{\rm min}) =
\int^{E_0 - E_e}_{\omega_{\rm min}}\frac{d\omega}{\omega}\frac{(E_0 -
  E_e - \omega)^2}{(E_0 - E_e)^2}\,\Big\{\Big(1 + \frac{\omega}{E_e} +
\frac{1}{2}\frac{\omega^2}{E^2_e}\Big) \,\Big[\frac{1}{\beta}\,{\ell
    n}\Big(\frac{1 + \beta}{1 - \beta}\Big) - 2\Big] +
\frac{\omega^2}{E^2_e}\Big\},
\end{eqnarray}
where $\omega_{\rm min}$ is a non-covariant infrared cut-off, which
can be also treated as the photon--energy threshold of the detector
\cite{Ivanov2013}.

We would like to emphasize that the function
$g^{(1)}_{\beta^-_c\gamma}(E_e, \mu)$ removes the dependence on the
infrared cut--off $\mu$ only in the neutron lifetime or in the
function $\zeta(E_e)$ \cite{Ivanov2013}. The dependence of the
radiative corrections on the infrared cut--off in the correlation
coefficients of the neutron beta decay should be removed by the
contributions of the corresponding correlation coefficients in the
neutron radiative beta decay \cite{Ivanov2013, Ivanov2017,
  Ivanov2019a}. The correlation coefficient of the neutron radiative
beta decay, caused by the correlation structure $(\vec{\xi}_n\cdot
\vec{k}_{\bar{\nu}})(\vec{\xi}_e \cdot \vec{k}_e)/E_e E_{\bar{\nu}}$,
we have calculated in section \ref{sec:appendix}. It is given by the
function $g^{(2)}_{\beta^-_c\gamma}(E_e, \mu)$ (see
Eqs.(\ref{eq:A.11}) and (\ref{eq:A.12})). Adding the contribution of
the neutron radiative beta decay to the correlation coefficient
$\zeta(E_e) T(E_e)$ in Eq.(\ref{eq:15}) with the definition of the
function $f_{\beta^-_c\gamma}(E_e, \mu)$ in Eq.(\ref{eq:16}) and
taking the limit $\mu \to 0$ we transcribe the correlation coefficient
$\zeta(E_e) T(E_e)$ into the form
\begin{eqnarray}\label{eq:19}
\hspace{-0.3in} \zeta(E_e) T(E_e) &=& - B_0\, \Big\{1 +
\frac{\alpha}{\pi}\,\Big[g_n(E_e) + \lim_{\mu \to 0}\Big(
  g^{(2)}_{\beta^-_c\gamma}(E_e, \mu) - g^{(1)}_{\beta^-_c\gamma}(E_e,
  \mu)\Big) + \frac{1 - \beta^2}{2\beta}\,{\ell n}\frac{1 + \beta}{1 -
    \beta}\Big] \Big\}\nonumber\\
\hspace{-0.3in}&& + \frac{1}{1 + 3 g^2_A}\,\frac{E_0}{m_N}\,\Big(2 g_A
\big(g_A + (\kappa + 1)\big) - \big(7g^2_A + g_A(3\kappa + 8) +
(\kappa + 1)\big)\, \frac{E_e}{E_0}\Big).
\end{eqnarray}
Since, according to \cite{Ivanov2013, Ivanov2017, Ivanov2019a}, the
difference $\lim_{\mu \to 0}\big(g^{(2)}_{\beta^-_c\gamma}(E_e,
\mu)\Big) - g^{(1)}_{\beta^-_c\gamma}(E_e, \mu)\big)$ is equal to
\begin{eqnarray}\label{eq:20}
\hspace{-0.3in} \lim_{\mu \to 0}\Big( g^{(2)}_{\beta^-_c\gamma}(E_e,
\mu) - g^{(1)}_{\beta^-_c\gamma}(E_e, \mu)\Big) =
\lim_{\omega_{\rm min} \to 0}\Big( g^{(2)}_{\beta^-_c\gamma}(E_e,
\omega_{\rm min}) - g^{(1)}_{\beta^-_c\gamma}(E_e, \omega_{\rm min})\Big),
\end{eqnarray}
we may rewrite Eq.(\ref{eq:19}) as follows
\begin{eqnarray}\label{eq:21}
\hspace{-0.3in} \zeta(E_e) T(E_e) &=& - B_0\, \Big\{1 +
\frac{\alpha}{\pi}\,\Big[g_n(E_e) + \lim_{\omega_{\rm min} \to 0}\Big(
  g^{(2)}_{\beta^-_c\gamma}(E_e,\omega_{\rm min}) -
  g^{(1)}_{\beta^-_c\gamma}(E_e, \omega_{\rm min})\Big) + \frac{1 -
    \beta^2}{2\beta}\,{\ell n}\frac{1 + \beta}{1 - \beta}\Big]
\Big\}\nonumber\\
\hspace{-0.3in}&& + \frac{1}{1 + 3 g^2_A}\,\frac{E_0}{m_N}\,\Big(2 g_A
\big(g_A + (\kappa + 1)\big) - \big(7 g^2_A + g_A(3 \kappa + 8) +
(\kappa + 1)\big)\, \frac{E_e}{E_0}\Big).
\end{eqnarray}
Using Eqs.(\ref{eq:18}) and (\ref{eq:A.11}), taking the limit
$\omega_{\rm min} \to 0$ and integrating over the photon energy
$\omega$ in the region $0 \le \omega \le E_0 - E_e$ (see, for example,
\cite{Ivanov2013}) we define the correlation coefficient $\zeta(E_e)
T(E_e)$ by taking into account the {\it outer} radiative corrections
of order $O(\alpha/\pi)$, calculated to leading order in the large
nucleon mass $m_N$ expansion, and the corrections of order
$O(E_e/m_N)$, caused by weak magnetism and proton recoil
\begin{eqnarray}\label{eq:22}
\hspace{-0.3in} \zeta(E_e) T(E_e) &=& - B_0\, \Big(1 +
\frac{\alpha}{\pi}\,g_n(E_e) + \frac{\alpha}{\pi}\, f_n(E_e) \Big) +
\frac{1}{1 + 3 g^2_A}\,\frac{E_0}{m_N}\,\Big(2 g_A \big(g_A + (\kappa
+ 1)\big) \nonumber\\
\hspace{-0.3in}&& - \big(7 g^2_A + g_A(3 \kappa + 8) + (\kappa +
1)\big)\, \frac{E_e}{E_0}\Big),
\end{eqnarray}
where the function $f_n(E_e)$ is equal to (see also \cite{Ivanov2013,
  Ivanov2017})
\begin{eqnarray}\label{eq:23}
\hspace{-0.3in}&& f_n(E_e) = \lim_{\omega_{\rm min} \to 0}\Big(
g^{(2)}_{\beta^-_c\gamma}(E_e, \omega_{\rm min}) -
g^{(1)}_{\beta^-_c\gamma}(E_e, \omega_{\rm min})\Big) + \frac{1 -
  \beta^2}{2\beta}\,{\ell n}\frac{1 + \beta}{1 - \beta} = \nonumber\\
\hspace{-0.3in}&&= \frac{1}{3}\,\frac{E_0 - E_e}{E_e}\Big(1 +
\frac{1}{8}\frac{E_0 - E_e}{E_e}\Big)\,\frac{1 - \beta^2}{\beta^2}\,
\Big[\frac{1}{\beta}\,{\ell n}\Big(\frac{1 + \beta}{1 - \beta}\Big) -
  2\Big]- \frac{1}{12}\frac{(E_0 - E_e)^2}{E^2_e} + \frac{1 -
  \beta^2}{2\beta}\,{\ell n}\Big(\frac{1 + \beta}{1 - \beta}\Big).
\end{eqnarray}
For the first time the function $f_n(E_e)$, defining the radiative
corrections $(\alpha/\pi)\,f_n(E_e)$ to the correlation coefficients
of the electron--antineutrino correlations $a(E_e)$ and of the
electron asymmetry $A(E_e)$, was calculated by Shann \cite{Shann1971}
(see also Eq.(D-58) in Appendix D in Ref. \cite{Ivanov2013}).

Using the correlation function $\zeta(E_e)$, calculated in
\cite{Ivanov2013} (see Eq.(6) with the replacement $\lambda = - g_A$),
we define the correlation coefficient $T(E_e)$
\begin{eqnarray}\label{eq:24}
\hspace{-0.3in}T(E_e) &=& - B_0\, \Big(1 + \frac{\alpha}{\pi}\,
f_n(E_e) \Big) + \frac{1}{1 + 3 g^2_A}\,\frac{E_0}{m_N}\,\Big(2 g_A
\big(g_A + (\kappa + 1)\big) - \big(7g^2_A + g_A(3 \kappa + 8) +
(\kappa + 1)\big)\, \frac{E_e}{E_0}\Big) \nonumber\\
\hspace{-0.3in}&+& \frac{B_0}{1 + 3 g^2_A}\, \frac{E_0}{m_N}\Big( - 2
g_A\big(g_A + (\kappa + 1)\big) + \big(10 g^2_A + 4 g_A (\kappa + 1) +
2\big)\, \frac{E_e}{E_0} - 2 g_A\big(g_A + (\kappa +
1)\big)\,\frac{m^2_e}{E^2_0}\,\frac{E_0}{E_e}\Big).
\end{eqnarray}
This is a complete description of the correlation coefficient
$T(E_e)$ at the level of $10^{-3}$ including the {\it outer}
radiative corrections of order $O(\alpha/\pi)$, calculated to leading
order in the large nucleon mass $m_N$ expansion, and the corrections
of order $O(E_e/m_N)$, caused by weak magnetism and proton recoil.

\section{Contributions of interactions beyond the Standard Model
  \cite{Jackson1957a}}
\label{sec:BSM}

For the calculation of the contributions of interactions beyond the SM
we use the effective phenomenological Lagrangian proposed in
\cite{Jackson1957a} (see also \cite{Ebel1957, Herczeg2001,
  Severijns2006}). It reads
\begin{eqnarray}\label{eq:25}
&&{\cal L}_{\rm BSM}(x) = - G_V \Big\{[\bar{\psi}_p(x) \gamma_{\mu}
    \psi_n(x)] [\bar{\psi}_e(x) \gamma^{\mu} (C_V + \bar{C}_V
    \gamma^5) \psi_{\nu_e}(x)] + [\bar{\psi}_p(x) \gamma_{\mu}
    \gamma^5 \psi_n(x)][\bar{\psi}_e(x) \gamma^{\mu} (\bar{C}_A + C_A
    \gamma^5) \psi_{\nu_e}(x)] \nonumber\\ && + [\bar{\psi}_p(x)
    \psi_n(x)][\bar{\psi}_e(x) (C_S + \bar{C}_S \gamma^5)
    \psi_{\nu_e}(x)] + [\bar{\psi}_p(x) \gamma ^5
    \psi_n(x)][\bar{\psi}_e(x) (C_P + \bar{C}_P \gamma^5)
    \psi_{\nu_e}(x)] + \frac{1}{2} [\bar{\psi}_p(x) \sigma^{\mu\nu}
    \gamma^5 \psi_n(x)] \nonumber\\ && \times [\bar{\psi}_e(x)
    \sigma_{\mu\nu} (\bar{C}_T + C_T \gamma^5) \psi_{\nu_e}(x) \Big\},
\end{eqnarray}
where we have followed the notations in \cite{Ivanov2013}.  The
effective phenomenological Lagrangian ${\cal L}_{\rm BSM}(x)$ reduces
to the standard effective Lagrangian ${\cal L}_W(x)$ of $V - A$ weak
low--energy interactions Eq.(\ref{eq:3}) by the replacement $C_V = -
\bar{C}_V = 1$, $C_A = - \bar{C}_A = g_A$ and $C_S = \bar{C}_S = C_P =
\bar{C}_P = C_T = \bar{C}_T = 0$. Following \cite{Ivanov2013} (see
Appendix G in Ref.\cite{Ivanov2013}) and skipping intermediate
calculations, carried out in the approximation of the leading order in
the large nucleon mass $m_N$ expansion, we get
\begin{eqnarray}\label{eq:26}
\hspace{-0.3in}\xi T(E_e) = - 2\, {\rm Re}\Big(C_V C^*_A +
\bar{C}_V \bar{C}^*_A + |C_A|^2 + |\bar{C}_A|^2 + C_S C^*_T +
\bar{C}_S \bar{C}^*_T - |C_T|^2 - |\bar{C}_T|^2\Big),
\end{eqnarray}
where the factor $\xi$ is equal to \cite{Jackson1957a, Ivanov2013}
\begin{eqnarray}\label{eq:27}
\hspace{-0.3in}\xi = |C_V|^2 + |\bar{C}_V|^2 + 3 |C_A|^2 +
3|\bar{C}_A|^2 + |C_S|^2 + |\bar{C}_S|^2 + 3|C_T|^2 + 3|\bar{C}_T|^2.
\end{eqnarray}
Our result in Eq.(\ref{eq:26}) agrees well with the result obtained by
Ebel and Feldman \cite{Ebel1957} but without contributions of
imaginary parts of the phenomenological scalar and tensor coupling
constants, which are proportional to the factor $\alpha m_e/k_e$,
caused by the distortion of the Dirac wave function of the electron in
the Coulomb field of the proton \cite{Jackson1957b}. In this
connections, we wold like to remind that, according to
\cite{Hardy2020, Severijns2019}, the phenomenological scalar coupling
constants should be zero, i.e. $C_S = \bar{C}_S = 0$. Then, we would
like to notice that the contribution of the phenomenological
pseudoscalar coupling constants $C_P$ and $\bar{C}_P$ vanishes to
leading order in the large nucleon mass expansion (see, for example,
\cite{Ivanov2020}).

At the replacement $C_V = - \bar{C}_V = 1$, $C_A = - \bar{C}_A = g_A$
and $C_S = \bar{C}_S = C_P = \bar{C}_P = C_T = \bar{C}_T = 0$ the
factor $\xi$ reduces to $\xi = 2(1 + 3 g^2_A)$. In the linear
approximation for the phenomenological vector and axial-vector
coupling constants $C_V = 1 + \delta C_V$, $\bar{C}_V = - 1 + \delta
\bar{C}_V$, $C_A = g_A + \delta C_A$ and $\bar{C}_A = - g_A + \delta
\bar{C}_A$ the contributions of the vector and axial-vector
phenomenological coupling constants can absorbed by renormalization of
the axial coupling constant $g_A$ and the Cabibbo-Kobayashi-Maskawa
(CKM) matrix element $V_{ud}$ (it is included into $G_V$)
\cite{Bhattacharya2012, Gardner2012, Cirigliano2012, Cirigliano2013a,
  Cirigliano2013b} (see also \cite{Ivanov2013, Ivanov2017,
  Ivanov2019a}). Thus, in the linear approximation for the
phenomenological vector, axial-vector, scalar and tensor coupling
constants \cite{Bhattacharya2012, Gardner2012, Cirigliano2012,
  Cirigliano2013a, Cirigliano2013b} (see also \cite{Ivanov2013,
  Ivanov2017, Ivanov2019a}) the contribution of interactions beyond
the SM, defined by the effective Lagrangian Eq.(\ref{eq:25}), is equal
to zero, i.e. $T^{(\rm BSM)}(E_e) = 0$.

\section{Contributions of the {\it second class} currents or the
  G--odd correlations}
\label{sec:Gparity}

The $G$--parity transformation, i.e. $G = C\,e^{\,i \pi I_2}$, where
$C$ and $I_2$ are the charge conjugation and isospin operators, was
introduced by Lee and Yang \cite{Lee1956a} as a symmetry of strong
interactions. According to the properties of hadronic currents under
$G$--transformation, Weinberg divided hadronic currents into two
classes, which are $G$--even {\it first class} and $G$--odd {\it
  second class} currents \cite{Weinberg1958}, respectively. Following
Weinberg \cite{Weinberg1958}, Gardner and Zhang \cite{Gardner2001},
and Gardner and Plaster \cite{Gardner2013} (see also \cite{Ivanov2017,
  Ivanov2019a}) the $G$--odd contributions to the matrix element of
the hadronic $n \to p$ transition, caused by the hadronic current, in
the $V - A$ theory of weak interactions can be taken in the form
\begin{eqnarray}\label{eq:28}
\langle
p(\vec{k}_p,\sigma_p)|J^{(+)}_{\mu}(0)|n(\vec{k}_n,\sigma_n)\rangle_{G
  - \rm odd} =
\bar{u}_p(\vec{k}_p,\sigma_p)\Big(\frac{q_{\mu}}{m_N}\,f_3(0) +
i\,\sigma_{\mu\nu}\gamma^5
\frac{q^{\nu}}{m_N}\,g_2(0)\Big)\,u_n(\vec{k}_n, \sigma_n),
\end{eqnarray}
where $J^{(+)}_{\mu}(0) = V^{(+)}_{\mu}(0) - A^{(+)}_{\mu}(0)$,
$\bar{u}_p(\vec{k}_p,\sigma_p)$ and $u_n(\vec{k}_n, \sigma_n)$ are the
Dirac wave functions of the proton and neutron
\cite{Ivanov2018}. Then, $f_3(0)$ and $g_2(0)$ are the
phenomenological coupling constants defining the strength of the {\it
  second class} currents in the weak decays. Following
\cite{Ivanov2018, Ivanov2019a} and skipping intermediate calculations
we get the contribution of the {\it second class} currents or the
$G$--odd correlations to the correlation coefficient $T(E_e)$
\begin{eqnarray}\label{eq:29}
T^{(\rm G-odd)}(E_e) = {\rm Re}g_2(0)\, \frac{1}{1 + 3 g^2_A}\,
\frac{E_0}{m_N}\, \Big(4 \big(g_A + 1\big) + 2 \big(g_A - 1\big)\,
\frac{E_e}{E_0}\Big).
\end{eqnarray}
Apart from the dependence of $T^{(\rm G-odd)}(E_e)$ on the axial
coupling constant $g_A$, the contribution of the {\it second class}
currents or the $G$--odd correlations to the correlation coefficient
$T(E_e)$ is represented by the phenomenological coupling constant
${\rm Re} g_2(0)$ only.

\section{Discussion}
\label{sec:Abschluss}

We have analyzed the correlation coefficient $T(E_e)$, caused by the
correlation structure $(\vec{\xi}_n \cdot
\vec{k}_{\bar{\nu}})(\vec{\xi}_e \cdot \vec{k}_e)/E_e E_{\bar{\nu}}$
invariant under discrete P, C and T symmetries. Such a correlation
structure was introduced by Ebel and Feldman \cite{Ebel1957} in
addition to the set correlation structures proposed by Jackson {\it et
  al.} \cite{Jackson1957a, Jackson1957b}.  The correlation coefficient
$T(E_e)$, calculated to leading order in the large nucleon mass $m_N$
expansion within the standard effective $V - A$ theory of weak
interactions \cite{Feynman1958}, is equal to $T(E_e) = - B_0$, where
$B_0 \sim 1$ \cite{Abele2008, Nico2009}. Having calculated the
correlation coefficient $T(E_e)$ to leading order in the large nucleon
mass $m_N$ expansion, we have given within the SM a complete
description of the correlation coefficient $T(E_e)$ at the level of
$10^{-3}$ by taking into account the {\it outer} model-independent
radiative corrections of order $O(\alpha/\pi)$, calculated to leading
order in the large nucleon mass expansion, and the corrections of
order $O(E_e/m_N)$, caused by weak magnetism and proton recoil. In
addition we have calculated the contributions of interactions beyond
the SM, expressed in terms of i) the phenomenological coupling
constants of the effective phenomenological interactions proposed by
Jackson {\it et al.}  \cite{Jackson1957a}, and ii) the
phenomenological coupling constants of the {\it second class}
currents, measuring the strength of $G$--odd correlations
\cite{Weinberg1958, Gardner2001, Gardner2013} (see also
\cite{Ivanov2018, Ivanov2019a}). We have found that in the linear
approximation for the phenomenological vector, axial-vector, scalar
and tensor coupling constants \cite{Bhattacharya2012, Gardner2012,
  Cirigliano2012, Cirigliano2013a, Cirigliano2013b} (see also
\cite{Ivanov2013, Ivanov2017, Ivanov2019a}) the contribution of
interactions beyond the SM, defined by the effective Lagrangian
Eq.(\ref{eq:25}), is equal to zero, i.e. $T^{(\rm BSM)}(E_e) = 0$. As
a result, in such an approximation the interactions beyond the SM are
represented in the correlation coefficient $T(E_e)$ by the {\it second
  class} currents or $G$--odd correlations only.  Our result in
Eq.(\ref{eq:26}) for the contribution of interactions beyond the SM,
defined by the phenomenological coupling constants of the
phenomenological interactions proposed by Jackson {\it et al.}
\cite{Jackson1957a}, agrees well with the result obtained by Ebel and
Feldman \cite{Ebel1957} but without contributions of the imaginary
parts of the phenomenological scalar and tensor coupling constants
proportional to the factor $\alpha m_e/k_e$, caused by the distortion
of the Dirac wave function of the electron in the Coulomb field of the
proton. Then, we would like to remind that, according to
\cite{Hardy2020, Severijns2019}, the contributions of the
phenomenological scalar coupling constants should vanish.

Summing up all corrections we obtain the following expression for the
correlation coefficient $T(E_e)$:
\begin{eqnarray}\label{eq:30}
\hspace{-0.3in}T(E_e) &=& - B_0\, \Big(1 + \frac{\alpha}{\pi}\,
f_n(E_e) \Big) + \frac{1}{1 + 3 g^2_A}\,\frac{E_0}{m_N}\,\Big(2 g_A
\big(g_A + (\kappa + 1)\big) - \big(7g^2_A + g_A(3 \kappa + 8) +
(\kappa + 1)\big)\, \frac{E_e}{E_0}\Big) \nonumber\\
\hspace{-0.3in}&+& \frac{B_0}{1 + 3 g^2_A}\, \frac{E_0}{m_N}\Big( - 2
g_A\big(g_A + (\kappa + 1)\big) + \big(10 g^2_A + 4 g_A (\kappa + 1) +
2\big)\, \frac{E_e}{E_0} - 2 g_A\big(g_A + (\kappa +
1)\big)\,\frac{m^2_e}{E^2_0}\,\frac{E_0}{E_e}\Big) \nonumber\\
\hspace{-0.3in}&+& {\rm Re}g_2(0)\, \frac{1}{1 + 3 g^2_A}\,
\frac{E_0}{m_N}\, \Big(4 \big(g_A + 1\big) + 2 \big(g_A - 1\big)\,
\frac{E_e}{E_0}\Big).
\end{eqnarray}
The correlation coefficient $T(E_e)$, calculated for $g_A =
1.2764$ \cite{Abele2018}, $E_0 = 1.2926\, {\rm MeV}$, $m_N =
938.9188\,{\rm MeV}$, $m_e = 0.5110\,{\rm MeV}$ and $\kappa = 3.7059$
\cite{PDG2020}, is equal to
\begin{eqnarray}\label{eq:31}
\hspace{-0.3in} T(E_e) &=& - \Big(0.987 - 4.63 \times 10^{-5} - 2.13
\times 10^{-3}\, {\rm Re} g_2(0)\Big) - 2.29 \times
10^{-3}\,f_n(E_e)\nonumber\\
\hspace{-0.3in}&& + \Big(2.94 \times 10^{-4} + 1.29 \times 10^{-4}\,
       {\rm Re} g_2(0)\Big)\, \frac{E_e}{E_0} - 5.51 \times 10^{-4} \,
       \frac{E_0}{E_e}.
\end{eqnarray}
According to \cite{Gardner2001, Gardner2013} and \cite{Ivanov2018,
  Ivanov2019a}, in the correlation coefficient $T(E_e)$ the
contribution of the {\it second class} currents may be estimated at
the level of a few parts of $10^{-5}$ and even smaller. Thus, the
numerical analysis of the correlation coefficient $T(E_e)$ shows that
such a correlation coefficient can be a nice tool for experimental
probes of contributions of the {\it second class} currents or $G$--odd
correlations in terms of the phenomenological coupling constant ${\rm
  Re}g_2(0)$ in Eq.(\ref{eq:28}). However, it is obvious that
successful experimental searches of such contributions it is required
the description of the correlation coefficient $T(E_e)$ within
the SM at the level of $10^{-5}$ and as well as experimental
uncertainties at the level of a few parts of $10^{-5}$. We are
planning to carry out such a theoretical description of the
correlation coefficient $T(E_e)$ in our forthcoming publication
by taking into account the results, obtained in \cite{Ivanov2019b,
  Ivanov2020a, Ivanov2020b} and also in \cite{Wilkinson1982} in terms
of Wilkinson's corrections (see also \cite{Ivanov2013, Ivanov2017,
  Ivanov2019a}).

A rather complicated correlation structure $(\vec{\xi}_n \cdot
\vec{k}_{\bar{\nu}})(\vec{\xi}_e \cdot \vec{k}_e)/E_e E_{\bar{\nu}}$
responsible for the correlation coefficient $T(E_e)$, which entangles
the spin--polarization vectors of the neutron and electron and
3-momenta of the electron and antineutrino, makes difficult its
experimental analysis. Indeed, the experimental investigation of the
correlation coefficient $T(E_e)$ should be performed for polarized
neutrons and longitudinally polarized decay electrons.  This is unlike
the experiments i) on the neutron beta decay for polarized neutrons
and transversally polarized electrons \cite{Bodek2019, Bodek2016,
  Kozela2009, Kozela2012} and ii) on the nuclear beta decays for
unpolarized nuclei and longitudinally polarized decay positrons
\cite{Severijns2011, Wichers1987, Carnoy1990, Carnoy1991}. Moreover,
the dependence of the correlation structure on the antineutrino
3--momentum demands a simultaneous detection of decay electrons and
protons, i.e.  electron-proton pairs, similar to the measurements of
the antineutrino asymmetry \cite{Serebrov1995, Serebrov1998,
  Schumann2007}. The theoretical analysis of the antineutrino
asymmetry in the neutron beta decay, related to the correlation
coefficient $B(E_e)$, was carried out by Gl\"uck {\it et a.}
\cite{Gluck1995} (see also \cite{Ivanov2013}). We are planning to
perform an analogous theoretical analysis of the asymmetry, related to
the correlation coefficient $T(E_e)$, in our forthcoming publication.

\section{Acknowledgements}

We thank Hartmut Abele for discussions stimulating this work. We thank
Vladimir Gudkov for calling our attention to the result obtained by
Ebel and Feldman \cite{Ebel1957}, which we have overlooked. The work
of A. N. Ivanov was supported by the Austrian ``Fonds zur F\"orderung
der Wissenschaftlichen Forschung'' (FWF) under contracts P31702-N27
and P26636-N20, and ``Deutsche F\"orderungsgemeinschaft'' (DFG) AB
128/5-2. The work of R. H\"ollwieser was supported by the Deutsche
Forschungsgemeinschaft in the SFB/TR 55. The work of M. Wellenzohn was
supported by the MA 23 (FH-Call 16) under the project ``Photonik -
Stiftungsprofessur f\"ur Lehre''.

\newpage

\section{\bf the Supplemental Material}
\label{sec:appendix}

\section*{Appendix A: The electron-photon-energy and angular
  distribution of the neutron radiative beta decay for a polarized
  neutron, a polarized electron and unpolarized proton and photon }
\renewcommand{\theequation}{A-\arabic{equation}}
\setcounter{equation}{0}

In order to remove the dependence of the radiative corrections to the
neutron lifetime and correlation coefficients of the neutron beta
decay on the infrared cut--off $\mu$ we have to add the contribution
of the neutron radiative beta decay $n \to p + e^- + \bar{\nu}_e +
\gamma$ \cite{Berman1958, Kinoshita1959} (see also \cite{Sirlin1967,
  Shann1971, Ando2004, Gudkov2006} and \cite{Ivanov2013, Ivanov2017,
  Ivanov2019a}). Following \cite{Ivanov2013, Ivanov2017, Ivanov2019a}
we define the electron-photon-energy and angular distribution of the
neutron radiative beta decay for a polarized neutron, a polarized
electron, a polarized photon and an unpolarized proton as follows
\begin{eqnarray}\label{eq:A.1}
\hspace{-0.3in}\frac{d^8\lambda_{\beta^-_c\gamma}(E_e,\omega,\vec{k}_e,
  \vec{k}_{\bar{\nu}},\vec{q},\vec{\xi}_n, \vec{\xi}_e)_{\lambda'
    \lambda}}{d\omega d E_e
  d\Omega_ed\Omega_{\bar{\nu}}\Omega_{\gamma}} &=& (1 + 3 g^2_A) \,
\frac{\alpha}{\pi}\,\frac{|G_V|^2}{(2\pi)^6}\,\sqrt{E^2_e -
  m^2_e}\,E_e\,F(E_e, Z = 1)\,(E_0 - E_e -
\omega)^2\,\frac{1}{\omega}\nonumber\\
\hspace{-0.3in}&&\times\,\sum_{\rm pol.}\frac{|M(n \to p e^-
  \bar{\nu}_e \gamma)|^2_{\lambda' \lambda} \omega^2}{(1 + 3 g^2_A) e^2|G_V|^2
  64 m^2_n E_e E_{\bar{\nu}}},
\end{eqnarray}
where we sum over polarizations of massive fermions. The photon state
is determined by the 4--momentum $q^{\mu} = (\omega, \vec{q}\,)$ and
the 4-vector of polarization $\varepsilon^{\mu}(q)_{\lambda}$ with
$\lambda = 1,2$, obeying the constraints
$\varepsilon^*(q)_{\lambda'}\cdot \varepsilon_{\lambda}(q) = -
\delta_{\lambda'\lambda}$ and $q \cdot \varepsilon_{\lambda}(q) =
0$. The sum over polarizations of the massive fermions is defined by
\cite{Ivanov2013, Ivanov2017, Ivanov2019a}
\begin{eqnarray}\label{eq:A.2}
\hspace{-0.15in}&&\sum_{\rm pol.}\frac{|M(n \to p e^- \bar{\nu}_e
  \gamma)|^2_{\lambda' \lambda} \omega^2}{(1 + 3 g^2_A) e^2|G_V|^2 64
  m^2_n E_e E_{\bar{\nu}}} = \frac{1}{(E_e - \vec{n}\cdot
  \vec{k}_e)^2}\, \frac{1}{(1 + 3 g^2_A) 32 E_e
  E_{\bar{\nu}}}\Big\{{\rm tr}\{(1 + \vec{\xi}_n\cdot
\vec{\sigma}\,)\}\,{\rm tr}\{(\hat{k}_e + m_e \gamma^5 \hat{\zeta}_e)
Q_{\lambda} \gamma^0 \hat{k}_{\bar{\nu}}\gamma^0
\bar{Q}_{\lambda'}\nonumber\\
\hspace{-0.15in}&&\times\, (1 - \gamma^5)\} + g_A {\rm tr}\{(1 +
\vec{\xi}_n\cdot \vec{\sigma}\,) \vec{\sigma}\,\}\cdot {\rm
  tr}\{(\hat{k}_e + m_e \gamma^5 \hat{\zeta}_e) Q_{\lambda} \gamma^0
\hat{k}_{\bar{\nu}}\vec{\gamma}\, \bar{Q}_{\lambda'} (1 - \gamma^5)\}
+ g_A {\rm tr}\{(1 + \vec{\xi}_n\cdot \vec{\sigma}\,)
\vec{\sigma}\,\}\cdot {\rm tr}\{(\hat{k}_e + m_e \gamma^5
\hat{\zeta}_e) \nonumber\\
\hspace{-0.15in}&&\times \, Q_{\lambda} \vec{\gamma}\,
\hat{k}_{\bar{\nu}}\gamma^0 \bar{Q}_{\lambda'} (1 - \gamma^5)\} +
g^2_A {\rm tr}\{(1 + \vec{\xi}_n\cdot \vec{\sigma}\,) \sigma^a\sigma^b
\} {\rm tr}\{(\hat{k}_e + m_e \gamma^5 \hat{\zeta}_e)
Q_{\lambda}\gamma^b \hat{k}_{\bar{\nu}}\gamma^a\, \bar{Q}_{\lambda'}
(1 - \gamma^5)\}\Big\},
\end{eqnarray}
where we have denoted a unit 3-vector $\vec{n} = \vec{q}/\omega$,
$Q_{\lambda} = 2\,\varepsilon^*_{\lambda}(q) \cdot k_e +
\hat{\varepsilon}^*_{\lambda}(q) \hat{q}$ and $\bar{Q}_{\lambda'} =
\gamma^0 Q^{\dagger}_{\lambda'} \gamma^0 = 2\,
\varepsilon_{\lambda'}(q) \cdot k_e +
\hat{q}\hat{\varepsilon}_{\lambda'}(q)$ \cite{Ivanov2013, Ivanov2017,
  Ivanov2019a}. Calculating the traces over the nucleon degrees of
freedom and using the properties of the Dirac matrices
\begin{eqnarray}\label{labelA.3}
\hspace{-0.3in}\gamma^{\alpha}\gamma^{\nu}\gamma^{\mu} =
\gamma^{\alpha}\eta^{\nu\mu} - \gamma^{\nu}\eta^{\mu\alpha} +
\gamma^{\mu}\eta^{\alpha\nu} +
i\,\varepsilon^{\alpha\nu\mu\beta}\,\gamma_{\beta}\gamma^5,
\end{eqnarray}
where $\eta^{\mu\nu}$ is the metric tensor of the Minkowski
space--time, $\varepsilon^{\alpha\nu\mu\beta}$ is the Levi--Civita
tensor defined by $\varepsilon^{0123} = 1$ and
$\varepsilon_{\alpha\nu\mu\beta}= - \varepsilon^{\alpha\nu\mu\beta}$
\cite{Itzykson1980}, we transcribe the right-hand-side (r.h.s.) of
Eq.(\ref{eq:A.2}) into the form \cite{Ivanov2013, Ivanov2017,
  Ivanov2019a}
\begin{eqnarray}\label{eq:A.4}
\hspace{-0.15in}&&\sum_{\rm pol.}\frac{|M(n \to p e^- \bar{\nu}_e
  \gamma)|^2_{\lambda' \lambda} \omega^2}{(1 + 3 g^2_A) e^2|G_V|^2 64
  m^2_n E_e E_{\bar{\nu}}} = \frac{1}{(E_e - \vec{n}\cdot
  \vec{k}_e)^2}\, \frac{1}{ 16 E_e}\Big\{\Big(1 + B_0\,
\frac{\vec{\xi}_n\cdot \vec{k}_{\bar{\nu}}}{E_{\bar{\nu}}}\Big){\rm
  tr}\{(\hat{k}_e + m_e \gamma^5 \hat{\zeta}_e) Q_{\lambda} \gamma^0
\bar{Q}_{\lambda'} (1 - \gamma^5)\} \nonumber\\
\hspace{-0.15in}&& + \Big(A_0\, \vec{\xi}_n +
a_0\,\frac{\vec{k}_{\bar{\nu}}}{E_{\bar{\nu}}}\Big) \cdot {\rm
  tr}\{(\hat{k}_e + m_e \gamma^5 \hat{\zeta}_e) Q_{\lambda}
\vec{\gamma}\, \bar{Q}_{\lambda'} (1 - \gamma^5)\}\Big\} = \frac{-
  B_0}{(E_e - \vec{n}\cdot \vec{k}_e)^2}\, \frac{\vec{\xi}_n\cdot
  \vec{k}_{\bar{\nu}}}{E_{\bar{\nu}}} \, \frac{m_e}{ 16 E_e}\,{\rm
  tr}\{\hat{\zeta}_e Q_{\lambda} \gamma^0 \bar{Q}_{\lambda'} (1 -
\gamma^5)\}\nonumber\\
\hspace{-0.15in}&& + \ldots \Big\},
\end{eqnarray}
where the ellipsis denotes the contributions of the terms, which
possess the correlation structures different to the correlation
structure responsible for the correlation coefficient $T(E_e)$.  In
Eq.(\ref{eq:A.4}) in the covariant form the traces over Dirac matrices
were calculated in \cite{Ivanov2013, Ivanov2019a}. The result is
\begin{eqnarray}\label{eq:A.5}
\hspace{-0.3in}&&\frac{1}{16}\,{\rm tr}\{\hat{\zeta}_e\, Q_{\lambda}
\gamma^{\mu} \bar{Q}_{\lambda'}(1 - \gamma^5)\} =
(\varepsilon^*_{\lambda}\cdot k_e)(\varepsilon_{\lambda'}\cdot k_e)
\zeta^{\mu}_e + \frac{1}{2}\,\Big( (\varepsilon^*_{\lambda}\cdot
k_e)(\varepsilon_{\lambda'}\cdot \zeta_e) +
(\varepsilon^*_{\lambda}\cdot \zeta_e)(\varepsilon_{\lambda'}\cdot
k_e) - (\varepsilon^*_{\lambda}\cdot \varepsilon_{\lambda'})
(\zeta_e\cdot q) \Big) q^{\mu} \nonumber\\
\hspace{-0.3in}&& - \frac{1}{2}\,\Big((\varepsilon^*_{\lambda}\cdot
k_e) \varepsilon^{\mu}_{\lambda'} + \varepsilon^{*\mu}_{\lambda}
(\varepsilon_{\lambda'}\cdot k_e)\Big) (\zeta_e\cdot q) -
i\,\frac{1}{2}\,\varepsilon^{\mu\nu\alpha\beta}\Big((\varepsilon^*_{\lambda}\cdot
k_e) \varepsilon_{\lambda' \nu} - \varepsilon^*_{\lambda
  \nu}(\varepsilon_{\lambda'}\cdot k_e)\Big) \zeta_{e \alpha}
q_{\beta} - i\,\frac{1}{2}\,q^{\mu}
\varepsilon^{\rho\varphi\alpha\beta} \varepsilon^*_{\lambda \rho}
\varepsilon_{\lambda' \varphi} \zeta_{e \alpha}q_{\beta}.\nonumber\\
\hspace{-0.3in}&&
\end{eqnarray}
The r.h.s. of Eq.(\ref{eq:A.5}) we calculate in the physical gauge
$\varepsilon_{\lambda} = (0, \vec{\varepsilon}_{\lambda})$
\cite{Ivanov2013, Ivanov2017, Ivanov2019a, Ivanov2017b}, where the
polarization vector $\vec{\varepsilon}_{\lambda}$ obeys the
constraints
\begin{eqnarray}\label{eq:A.6}
\hspace{-0.3in}\vec{q}\cdot \vec{\varepsilon}^{\,*}_{\lambda} &=&
\vec{q}\cdot \vec{\varepsilon}_{\lambda'} =
0\;,\;\vec{\varepsilon}^{\,*}_{\lambda}\cdot
\vec{\varepsilon}_{\lambda'} = \delta_{\lambda \lambda'} \;,\;
\sum_{\lambda = 1,2}\vec{\varepsilon}^{\,i
  *}_{\lambda}\vec{\varepsilon}^{\,j}_{\lambda} = \delta^{ij} -
\frac{\vec{k}^{\,i} \vec{k}^{\,j}}{\omega^2} = \delta^{ij} -
\vec{n}^{\,i}\, \vec{n}^{\,j}\;,\;\sum_{j =1,2,3}\sum_{\lambda =
  1,2}\vec{\varepsilon}^{\,j
  *}_{\lambda}\vec{\varepsilon}^{\,j}_{\lambda} = 2.
\end{eqnarray}
In the physical gauge $\varepsilon_{\lambda} = (0,
\vec{\varepsilon}_{\lambda})$ we obtain for the r.h.s. of
Eq.(\ref{eq:A.5}) the following expression
\begin{eqnarray}\label{eq:A.7}
\hspace{-0.15in}\sum_{\rm pol.}\frac{|M(n \to p e^- \bar{\nu}_e
  \gamma)|^2_{\lambda' \lambda} \omega^2}{(1 + 3 g^2_A) e^2|G_V|^2 16
  m^2_n E_e E_{\bar{\nu}}} &=& \frac{- B_0}{(E_e - \vec{n}\cdot
  \vec{k}_e)^2} \frac{m_e}{E_e} \Big\{
(\vec{\varepsilon}^{\,*}_{\lambda}\cdot
\vec{k}_e)(\vec{\varepsilon}_{\lambda'}\cdot \vec{k}_e) \zeta^0_e +
\frac{1}{2}\,\Big((\vec{\varepsilon}^{\,*}_{\lambda}\cdot
\vec{k}_e)(\vec{\varepsilon}_{\lambda'}\cdot \vec{\zeta}_e) \nonumber\\
\hspace{-0.15in}&& +
\frac{1}{2}\,(\vec{\varepsilon}^{\,*}_{\lambda}\cdot
\vec{\zeta}_e)(\vec{\varepsilon}_{\lambda'}\cdot \vec{k}_e)\Big)
\omega + (\vec{\varepsilon}^{\,*}_{\lambda}\cdot
\vec{\varepsilon}_{\lambda'})\big(\zeta^0_e \omega - \vec{\zeta}_e
\cdot \vec{q}\,\big)\, \omega + \ldots\Big\},
\end{eqnarray}
where the ellipsis denotes the contribution of the terms, which vanish
after the summation over polarizations of the photon. Plugging
Eq.(\ref{eq:A.7}) into Eq.(\ref{eq:A.1}) we obtain the contribution to
the electron-photon-energy and angular distribution of the neutron
radiative beta decay for a polarized neutron, a polarized electron, a
polarized photon and an unpolarized proton, which should be
responsible for a cancellation of the dependence of the radiative
corrections to the correlation coefficient $T(E_e)$ on the infrared
cut--off. We get
\begin{eqnarray}\label{eq:A.8}
\hspace{-0.3in}&&\frac{d^8\lambda_{\beta^-_c\gamma}(E_e,\omega,\vec{k}_e,
  \vec{k}_{\bar{\nu}},\vec{q},\vec{\xi}_n, \vec{\xi}_e)_{\lambda'
    \lambda}}{d\omega d E_e
  d\Omega_ed\Omega_{\bar{\nu}}\Omega_{\gamma}} = (1 + 3 g^2_A) \,
\frac{\alpha}{\pi}\,\frac{|G_V|^2}{(2\pi)^6}\,\sqrt{E^2_e -
  m^2_e}\,E_e\,F(E_e, Z = 1)\,(E_0 - E_e -
\omega)^2\,\frac{1}{\omega}\nonumber\\
\hspace{-0.3in}&&\times\,\bigg\{- B_0\,\frac{\vec{\xi}_n \cdot
  \vec{k}_{\bar{\nu}}}{E_{\bar{\nu}}} \frac{m_e}{E_e} \frac{1}{(E_e -
  \vec{n}\cdot
  \vec{k}_e)^2}\,\Big((\vec{\varepsilon}^{\,*}_{\lambda}\cdot
\vec{k}_e)(\vec{\varepsilon}_{\lambda'}\cdot \vec{k}_e) \zeta^0_e +
\frac{1}{2}\,\Big((\vec{\varepsilon}^{\,*}_{\lambda}\cdot
\vec{k}_e)(\vec{\varepsilon}_{\lambda'}\cdot \vec{\zeta}_e) +
\frac{1}{2}\,(\vec{\varepsilon}^{\,*}_{\lambda}\cdot
\vec{\zeta}_e)(\vec{\varepsilon}_{\lambda'}\cdot \vec{k}_e)\Big)
\omega \nonumber\\
\hspace{-0.15in}&&+ \frac{1}{2}\,(\vec{\varepsilon}^{\,*}_{\lambda}\cdot
\vec{\varepsilon}_{\lambda'})\big(\zeta^0_e \omega - \vec{\zeta}_e
\cdot \vec{q}\,\big)\, \omega + \ldots\Big) + \ldots\bigg\},
\end{eqnarray}
where ellipses denote the contributions of the terms, which vanish
after the summation over polarizations of the photon and which define
the contributions of other correlation structures \cite{Ivanov2013,
  Ivanov2017, Ivanov2019a}, respectively. Summing up over
polarizations of the photon we determine the electron-photon-energy
and angular distribution of the neutron radiative beta decay in terms
of the integrals over directions of the 3--momentum of the photon
\begin{eqnarray}\label{eq:A.9}
\hspace{-0.3in}&&\frac{d^6\lambda_{\beta^-_c\gamma}(E_e,\omega,\vec{k}_e,
  \vec{k}_{\bar{\nu}},\vec{\xi}_n, \vec{\xi}_e)}{d\omega d E_e
  d\Omega_ed\Omega_{\bar{\nu}}} = (1 + 3 g^2_A) \,
\frac{\alpha}{\pi}\,\frac{|G_V|^2}{16\pi^5}\,\sqrt{E^2_e -
  m^2_e}\,E_e\,F(E_e, Z = 1)\,(E_0 - E_e - \omega)^2\,
\frac{1}{\omega}\nonumber\\
\hspace{-0.3in}&&\times\,\bigg\{- B_0\,\frac{\vec{\xi}_n \cdot
  \vec{k}_{\bar{\nu}}}{E_{\bar{\nu}}}\,\frac{m_e}{E_e}\int
\frac{d\Omega_{\gamma}}{4\pi}\Big\{\zeta^0_e\, \frac{k^2_e -
  (\vec{n}\cdot \vec{k}_e)^2}{(E_e - \vec{n}\cdot \vec{k}_e)^2} +
\omega\, \frac{(\vec{k}_e \cdot \vec{\zeta}_e) - (\vec{n}\cdot
  \vec{k}_e)(\vec{n}\cdot \vec{\zeta}_e)}{(E_e - \vec{n}\cdot
  \vec{k}_e)^2} + \omega^2 \, \frac{\zeta^0_e - \vec{n}\cdot
  \vec{\zeta}_e}{(E_e - \vec{n}\cdot \vec{k}_e)^2}\Big\} +
\ldots\bigg\}.
\end{eqnarray}
For the calculation of the integrals over the directions of the
3--momentum of the photon we use the results, obtained in
\cite{Ivanov2017}. Having integrated over the directions of the
3-momentum and energy of the photon we get
\begin{eqnarray}\label{eq:A.10}
\hspace{-0.3in}&&\frac{d^6\lambda_{\beta^-_c\gamma}(E_e, \vec{k}_e,
  \vec{k}_{\bar{\nu}},\vec{\xi}_n, \vec{\xi}_e)}{d E_e
  d\Omega_ed\Omega_{\bar{\nu}}} = (1 + 3 g^2_A) \,
\frac{\alpha}{\pi}\,\frac{|G_V|^2}{16\pi^5}\,\sqrt{E^2_e -
  m^2_e}\,E_e\,F(E_e, Z = 1)\,(E_0 - E_e)^2\nonumber\\
\hspace{-0.3in}&&\times\,\bigg\{- B_0\,\frac{(\vec{\xi}_n \cdot
  \vec{k}_{\bar{\nu}})(\vec{\xi}_e \cdot \vec{k}_e)}{E_e
  E_{\bar{\nu}}}\,g^{(2)}_{\beta^-_c\gamma}(E_e, \omega_{\rm min}) +
\ldots\Big\},
\end{eqnarray}
where the function $g^{(2)}_{\beta^-_c\gamma}(E_e, \omega_{\rm min})$
is defined by the integral
\begin{eqnarray}\label{eq:A.11}
\hspace{-0.3in}g^{(2)}_{\beta^-_c\gamma}(E_e,\omega_{\rm min}) =
\int^{E_0 - E_e}_{\omega_{\rm min}}\frac{d\omega}{\omega}\,\frac{(E_0
  - E_e - \omega)^2}{(E_0 - E_e)^2}\,\Big[1 +
  \frac{1}{\beta^2}\frac{\omega}{E_e}\Big(1 +
  \frac{1}{2}\frac{\omega}{E_e}\Big)\Big]\,\Big[\frac{1}{\beta}\,{\ell
    n}\Big(\frac{1 + \beta}{1 - \beta}\Big) - 2\Big].
\end{eqnarray}
The function $g^{(2)}_{\beta^-_c\gamma}(E_e,\mu)$, regularized by the
covariant infrared cut--off $\mu$, was calculated in \cite{Ivanov2013}
(see Eq.(B-28) in Appendix B in Ref.\cite{Ivanov2013}). We adduce the
functions $g^{(1)}(E_e, \mu)$ and $g^{(2)}_{\beta^-_c\gamma}(E_e,\mu)$
for completeness
\begin{eqnarray}\label{eq:A.12}
\hspace{-0.3in}&&g^{(1)}_{\beta^-_c\gamma}(E_e,\mu) = \Big[{\ell
    n}\Big(\frac{2(E_0 - E_e)}{\mu}\Big) - \frac{3}{2} +
  \frac{1}{3}\,\frac{E_0 - E_e}{E_e}\, \Big(1 + \frac{1}{8} \frac{E_0
    - E_e}{E_e} \Big)\Big]\Big[\frac{1}{\beta}\,{\ell n}\Big(\frac{1 +
    \beta}{1 - \beta}\Big) - 2\Big]\nonumber\\
\hspace{-0.3in}&& + 1 + \frac{1}{2\beta}\,{\ell n}\Big(\frac{1 +
  \beta}{1 - \beta}\Big) - \frac{1}{4\beta}\,{\ell n}^2\Big(\frac{1 +
  \beta}{1 - \beta}\Big) - \frac{1}{\beta}\,{\rm Li}_2\Big(\frac{2
  \beta}{1 + \beta} \Big) + \frac{1}{12} \frac{(E_0 -
  E_e)^2}{E^2_e},\nonumber\\
\hspace{-0.3in}&&g^{(2)}_{\beta^-_c\gamma}(E_e,\mu) = \Big[{\ell
    n}\Big(\frac{2(E_0 - E_e)}{\mu}\Big) - \frac{3}{2} +
  \frac{1}{3}\,\frac{E_0 - E_e}{\beta^2 E_e}\, \Big(1 +
  \frac{1}{8}\,\frac{E_0 - E_e}{E_e}\Big)
  \Big]\Big[\frac{1}{\beta}\,{\ell n}\Big(\frac{1 + \beta}{1 -
    \beta}\Big) - 2\Big]\nonumber\\
\hspace{-0.3in}&& + 1 + \frac{1}{2\beta}\,{\ell n}\Big(\frac{1 +
  \beta}{1 - \beta}\Big) - \frac{1}{4\beta}\,{\ell n}^2\Big(\frac{1 +
  \beta}{1 - \beta}\Big) - \frac{1}{\beta}\,{\rm Li}_2\Big(\frac{2
  \beta}{1 + \beta} \Big),
\end{eqnarray}
where ${\rm Li}_2(z)$ is the PolyLogarithmic function
\cite{Mitchell1949}. Taking into account the contribution of the
neutron radiative beta decay, given by Eqs.(\ref{eq:A.12}) and
(\ref{eq:A.11}), we remove the dependence of the radiative corrections
to the correlation coefficient $\zeta(E_e) T(E_e)$ on the infrared
cut-off (see Eqs.(\ref{eq:19}) - (\ref{eq:23})).


\newpage


\begin{thebibliography}{9}
\bibitem{Chadwick1932} J. Chadwick, {\it Possible existence of a
  neutron}, Nature {\bf 129}, 312 (1932); \\ DOI: 10.1038/129312a0.

\bibitem{Abele2008} H. Abele, {\it The neutron. Its properties and
  basic interactions}, Progr. Part. Nucl. Phys. {\bf 60}, 1 (2008);
  \\ DOI: https://doi.org/10.1016/j.ppnp.2007.05.002.

\bibitem{Nico2009} J. S. Nico, {\it Neutron beta decay}, J. Phys. G:
  Nucl. Part. Phys. {\bf 36}, 104001 (2009); \\ DOI:
  https://doi.org/10.1088/0954-3899/36/10/104001.
  
\bibitem{Serebrov2007} A. P. Serebrov, {\it Neutron
  $\beta$--decay. Standard Model and Cosmology}, Phys. Lett. B {\bf
  650}, 321 (2007); \\ DOI:
  https://doi.org/10.1016/j.physletb.2007.05.047.

\bibitem{Paul2009} S. Paul, {\it The puzzle of neutron lifetime},
  Nucl. Instrum. Meth. A {\bf 611}, 157 (2009); \\ DOI:
  10.1016/j.nima.2009.07.095.
  
\bibitem{Dubbers2011} D. Dubbers and M. G. Schmid, {\it The neutron
  and its role in cosmology and particle physics}, Rev. Mod. Phys. {\bf
    83}, 1111 (2011); \\ DOI:
  https://doi.org/10.1103/RevModPhys.83.1111.

\bibitem{Paul2012} St. Paul, {\it The neutron and the Universe -
  History of a relationship}, PoS BORMIO2012, 025 (2012); \\ DOI:
  https://doi.org/10.22323/1.160.0025.
  
\bibitem{Serebrov2016} A. V. Chechkin, A. V. Ivanchik, A. P. Serebrov,
  and S.V. Bobashev, {\it Effect of the neutron lifetime on processes
    in the early universe}, Tech. Phys. {\bf 61}, 1101 (2016); \\ DOI:
  https://doi.org/10.1134/S1063784216070069.

\bibitem{Lee1956} T. D. Lee and C. N. Yang, {\it Question of parity
  conservation in weak interactions}, Phys. Rev. {\bf 104}, 254
  (1956); \\ DOI: https://doi.org/10.1103/PhysRev.104.254.
  

\bibitem{Lee1957a} T. D. Lee and C. N. Yang, {\it Parity
  nonconservation and a two-component theory of the neutrino},
  Phys. Rev. {\bf 105}, 1671 (1957); \\ DOI:
  https://doi.org/10.1103/PhysRev.105.1671.

\bibitem{Lee1957b} T. D. Lee, R. Oehme, and C. N. Yang, {\it Remarks
  on possible noninvariance under time reversal and charge
  conjugation}, Phys. Rev. {\bf 106}, 340 (1957); \\ DOI:
  https://doi.org/10.1103/PhysRev.106.340.

\bibitem{Severijns2011} N. Severijns and O. Naviliat-Cuncic, {\it
  Symmetry tests in nuclear beta decay},
  Annu. Rev. Nucl. Part. Sci. {\bf 61}, 23 (2011); \\ DOI:
  https://doi.org/10.1146/annurev-nucl-102010-130410.
  
\bibitem{DGH2014} J. F. Gonoghue, E. Golowich, and B. R. Holstein, in
  {\it Dynamics of the Standard Model}, 2nd edition, Cambridge
  University Press, Cambridge 2014; \\ DOI:
  https://doi.org/10.1017/CBO9780511803512.

\bibitem{PDG2020} P. A. Zyla {\it et al.},{\it Review of particle
  physics} (Particle Data Group), Prog. Theor. Exp. Phys. {\bf 2020},
  083C01 (2020); \\ DOI: https://doi.org/10.1093/ptep/ptaa104.
    
 

\bibitem{Abele2016} H. Abele, {\it Precision experiments with cold and
  ultra-cold neutrons}, Hyperfine Interact. {\bf 237}, 155 (2016);
  \\ DOI: https://doi.org/10.1007/s10751-016-1352-z.


\bibitem{Serebrov2017} A. P. Serebrov, {\it Research of fundamental
  interactions with use of ultracold neutrons},
  J. Phys. Conf. Ser. {\bf 798}, 012206 (2017); \\ DOI:
  https://doi.org/10.1088/1742-6596/798/1/012206.

\bibitem{Bodek2019} K. Bodek, L. De Keukeleere, M. Kolodziej,
  A. Kozela, M. Kuzniak, K. Lojek, M. Perkowski, H. Przybilski,
  K. Pysz, D. Rozpedzik, N. Severijns, T. Soldner, A. R. Young, and
  J. Zejma, {\it BRAND – Search for BSM physics at TeV scale by
    exploring transverse polarization of electrons emitted in neutron
    decay}, International Workshop on Particle Physics at Neutron
  Sources 2018 (PPNS 2018), EPJ Web of Conferences {\bf 219}, 04001
  (2019); \\ DOI: https://doi.org/10.1051/epjconf/201921904001.

\bibitem{Bhattacharya2012} T. Bhattacharya, V. Cirigliano,
  S. D. Cohen, A. Filipuzzi, M. Gonz\'alez-Alonso, M. L. Graesser,
  R. Gupta, and Huey-Wen Lin, {\it Probing novel scalar and tensor
    interactions from (ultra)cold neutrons to the LHC}, Phys. Rev. D
  {\bf 85}, 054512 (2012); \\ DOI:
  https://doi.org/10.1103/PhysRevD.85.054512.

\bibitem{Gardner2012} S. Gardner, V. Cirigliano, P. Fierlinger,
  C. Fischer, K. Jansen, S. Paul, F. J. Llanes-Estrada, H. W. Lin, and
  W. M. Snow, {\it Round Table: Resolving Physics BSM at Low
    Energies}, PoS ConfinementX (2012) 024; \\ DOI:
  https://doi.org/10.22323/1.171.0024.

\bibitem{Cirigliano2012} V. Cirigliano, M. Gonz\'alez-Alonso,
M. L. Graesser, {\it Non-standard charged current interactions: beta
  decays versus the LHC}, JHEP {\bf 2012}, 25 (2012); \\ DOI:
https://doi.org/10.1007/JHEP10(2012)025.

\bibitem{Cirigliano2013a} V. Cirigliano and M. J. Ramsey-Musolf, {\it
  Low energy probes of physics beyond the Standard Model},
  Prog. Part. Nucl. Phys. {\bf 71}, 2 (2013); \\ DOI:
  https://doi.org/10.1016/j.ppnp.2013.03.002.

\bibitem{Cirigliano2013b} V. Cirigliano, S. Gardner, and B. Holstein,
  {\it Beta decays and non-standard interactions in the LHC era},
  Prog. Part. Nucl. Phys. {\bf 71}, 93 (2013); \\ DOI:
  https://doi.org/10.1016/j.ppnp.2013.03.005.
  
\bibitem{Severijns2015} N. Severijns, {\it Low-energy weak interaction
  physics in the LHC era}, JPS Conf. Proc. {\bf 6}, 010003 (2015);
  \\ DOI: https://doi.org/10.7566/JPSCP.6.010003.

\bibitem{Bodek2016} K. Bodek, {\it Beta-decay correlations in the LHC
  era}, Acta Phys. Polon. B {\bf 47}, 349 (2016); \\ DOI:
  10.5506/APhysPolB.47.349.

  
\bibitem{Jackson1957a} J. D. Jackson, S. B. Treiman, and H. W. Wyld
  Jr., {\it Possible tests of time reversal invariance in beta decay},
  Phys. Rev. {\bf 106}, 517 (1957); \\ DOI:
  https://doi.org/10.1103/PhysRev.106.517.

\bibitem{Jackson1957b} J. D. Jackson, S. B. Treiman, and H. W. Wyld
  Jr., {\it Coulomb corrections in allowed beta transitions},
  Nucl. Phys. {\bf 4}, 206 (1957); DOI:
  https://doi.org/10.1016/0029-5582(87)90019-8.


\bibitem{Jackson1958} J. D. Jackson, S. B. Treiman, and H. W. Wyld,
  Jr., {\it Note on relativistic coulomb wave functions},
  Z. Phys. {\bf 150}, 640 (1958); DOI:
  https://doi.org/10.1007/BF01340460.
  

\bibitem{Ivanov2013} A. N. Ivanov, M. Pitschmann, and
  N. I. Troitskaya, {\it Neutron $\beta$--decay as a laboratory for
    testing the standard model}, Phys. Rev. D {\bf 88}, 073002 (2013);
  \\ DOI: https://doi.org/10.1103/PhysRevD.88.073002; arXiv:1212.0332
     [hep--ph].

 \bibitem{Ivanov2017} A. N. Ivanov, R. H\"ollwieser, N. I. Troitskaya,
   M. Wellenzohn, and Ya. A. Berdnikov, {\it Precision analysis of
     electron energy spectrum and angular distribution of neutron beta
     decay with polarized neutron and electron}, Phys. Rev. C {\bf
     95}, 055502 (2017); \\ DOI: 10.1103/PhysRevC.95.055502;
   arXiv:1705.07330 [hep-ph].

\bibitem{Ivanov2018} A. N. Ivanov, R. H\"ollwieser, N. I. Troitskaya,
  M. Wellenzohn, and Ya. A. Berdnikov, {\it Tests of the standard
    model in neutron beta decay with polarized neutron and electron
    and an unpolarized proton}, Phys. Rev. D 98, 035503 (2018);
  \\ DOI: 10.1103/PhysRevD.99.053004; arXiv:1805.03880 [hep-ph].

\bibitem{Ivanov2019a} A. N. Ivanov, R. H\"ollwieser, N. I. Troitskaya,
  M. Wellenzohn, and Ya. A. Berdnikov, {\it Test of the Standard Model
    in neutron beta decay with polarized electrons and unpolarized
    neutrons and protons}, Phys. Rev. D 99, 053004 (2019); \\ DOI:
  10.1103/PhysRevD.99.053004; arXiv:1811.04853 [hep-ph].

 \bibitem{Ivanov2019y} A. N. Ivanov, R. H\"ollwieser, N. I. Troitskaya,
  M. Wellenzohn, and Ya. A. Berdnikov, {\it Neutron dark matter decays
    and correlation coefficients of neutron beta decays},
  Nucl. Phys. B {\bf 938}, 114 (2019); \\ DOI:
  https://doi.org/10.1016/j.nuclphysb.2018.11.005.

 
\bibitem{Ivanov2019b} A. N. Ivanov, R. H\"ollwieser, N. I. Troitskaya,
  M. Wellenzohn, and Ya. A. Berdnikov, {\it Radiative corrections of
    order $O(\alpha E_e/m_N)$ to Sirlin's radiative corrections of
    order $O(\alpha/\pi)$ to the neutron lifetime}, Phys. Rev. D {\bf
    99}, 093006 (2019);\\ DOI:
  https://doi.org/10.1103/PhysRevD.99.093006; arXiv:1905.01178
  [hep-ph].

\bibitem{Ivanov2020} A. N. Ivanov, R. H\"ollwieser, N. I. Troitskaya,
  M. Wellenzohn, and Ya. A. Berdnikov, {\it Precision analysis of
    pseudoscalar interactions in neutron beta decays}, Nucl. Phys. B
  {\bf 951}, 114891 (2020); \\ DOI:
  https://doi.org/10.1016/j.nuclphysb.2019.114891; arXiv:1905.04147
  [hep-ph].
  
\bibitem{Ivanov2020a} A. N. Ivanov, R. H\"ollwieser, N. I. Troitskaya,
  M. Wellenzohn, and Ya. A. Berdnikov, {\it Radiative corrections of
    order $O(\alpha E_e/m_N)$ to Sirlin's radiative corrections of
    order $O(\alpha/\pi)$, induced by the hadronic structure of the
    neutron , to the neutron lifetime}




\bibitem{Abele2018} B. M\"arkisch, H. Mest, H. Saul, X. Wang,
  H. Abele, D. Dubbers, M. Klopf, A. Petoukhov, C. Roick, T. Soldner,
  and D. Werder, {\it Measurement of the weak axial-vector coupling
    constant in the decay of free neutrons using a pulsed cold neutron
    beam}, Phys. Rev. Lett. {\bf 122}, 242501 (2019); \\ DOI:
  https://doi.org/10.1103/PhysRevLett.122.242501; arXiv: 1812.04666
  [nucl-ex].

\bibitem{Sirlin2018} A. Czarnecki, W. J. Marciano, and A. Sirlin, {\it
  Neutron lifetime and axial coupling constant}, Phys. Rev. Lett. {\bf
  120}, 202002 (2018); \\ DOI:
  https://doi.org/10.1103/PhysRevLett.120.202002; arXiv: 1802.01804
  [hep-ph].
  

\bibitem{Ivanov2020b} A. N. Ivanov, R. H\"ollwieser, N. I. Troitskaya,
  M. Wellenzohn, and Ya. A. Berdnikov, {\it Corrections of order
    $O(E^2_e/m^2_N)$, caused by weak magnetism and proton recoil, to
    the neutron lifetime and correlation coefficients of the neutron
    beta decay}, Results in Physics {\bf 21}, 103806 (2021); \\ DOI:
  https://doi.org/10.1016/j.rinp.2020.103806; arXiv: 2010.14336
  [hep-ph].

 \bibitem{Itzykson1980} C. Itzykson and J.--B. Zuber, in {\it Quantum
  field theory}, McGraw--Hill Inc., New York, 1980.
 
  

\bibitem{Blatt1952} 
J. M. Blatt and V. F. Weisskopf,  {\it Theoretical nuclear physics},
John Wily $\&$ Sons, New York 1952.

  
\bibitem{Wilkinson1982} D. H. Wilkinson, {\it Analysis of neutron beta
  decay}, Nucl. Phys. A {\bf 377}, 474 (1982); \\ DOI:
  https://doi.org/10.1016/0375-9474(82)90051-3.

 \bibitem{Antognini2013} A.  Antognini {\it et al.},
  {\it Proton structure from the measurement of 2S-2P transition
    frequencies of muonic hydrogen}, Science {\bf 339} 417 (2013);
  \\ DOI: 10.1126/science.1230016.

\bibitem{Fierz1937} M. Fierz, {\it Zur Fermischen Theorie des
  $\beta$-Zerfalls}, Z. Physik {\bf 104}, 553 (1937); \\ DOI:
  https://doi.org/10.1007/BF01330070.

  
\bibitem{Hardy2020} J. C. Hardy and I. S. Towner, {\it Superallowed
  $0^+ \to 0^+$ nuclear beta decays: 2020 critical survey, with
  implications for $V_{ud}$ and CKM unitarity}, Phys. Rev. C {\bf
  102}, 045501 (2020); \\ DOI:
  https://doi.org/10.1103/PhysRevC.102.045501.

\bibitem{Severijns2019} M. Gonz\'alez--Alonso, O. Naviliat--Cuncic,
  and N. Severijns, {\it New physics searches in nuclear and neutron
    beta decay}, Prog. Part. Nucl. Phys. {\bf 104}, 165 (2019);
  \\ DOI: https://doi.org/10.1016/j.ppnp.2018.08.002.

\bibitem{Abele2019} H. Saul, Ch. Roick, H. Abele, H. Mest, M. Klopf,
  A. Petukhov, T. Soldner, X. Wang, D. Werder, and B.  M\"arkisch,
  {\it Limit on the Fierz interference term b from a measurement of
    the beta asymmetry in neutron decay}, Phys. Rev. Lett. {\bf 125},
  112501 (2020); \\ DOI:
  https://doi.org/10.1103/PhysRevLett.125.112501.

\bibitem{Young2019} V. Cirigliano, A. Garcia, D. Gazit,
  O. Naviliat-Cuncic, G. Savard, and A. Young, {\it Precision beta
    decay as a probe of new physics}, arXiv:1907.02164 [nucl-ex].

\bibitem{Sun2020} X. Sun {\it et al.}, {\it Improved limits on Fierz
  interference using asymmetry measurements from the ultracold neutron
  asymmetry (UCNA) experiment} (UCNA Collaboration), Phys.  Rev. C
  {\bf 101}, 035503 (2020); \\ DOI:
  https://doi.org/10.1103/PhysRevC.101.035503.


 \bibitem{Bilenky1959} S. M. Bilen'kii, R. M. Ryndin,
  Ya. A. Smorodinskii, and Ho Tso-Hsiu, {\it On the theory of the neutron
  beta decay}, JETP {\bf 37}, 1759 (1959) (in Russian);
  Sov. Phys. JETP, {\bf 37}(10), 1241 (1960). 

\bibitem{Ando2004} S. Ando, H. W. Fearing, V. Gudkov, K. Kubodera,
  F. Myhrer, S. Nakamura, and T. Sato, {\it Neutron beta--decay in
    effective field theory}, Phys. Lett. B {\bf 595}, 250 (2004); DOI:
  https://doi.org/10.1016/j.physletb.2004.06.037.
  
\bibitem{Gudkov2006} V. Gudkov, G. I. Greene, and J. R. Calarco, {\it
  General classification and analysis of neutron beta-decay
  experiments}, Phys. Rev. C {\bf 73}, 035501 (2006); \\ DOI:
  https://doi.org/10.1103/PhysRevC.73.035501.


\bibitem{Callan1967} C. G. Callan and S. B. Treiman, {\it Electromagnetic
  simulation of T violation in beta decay}, Phys. Rev. {\bf 162}, 1494
  (1967); \\ DOI: https://doi.org/10.1103/PhysRev.162.1494.
  
\bibitem{Ando2009} S. I. Ando, J. A. McGovern, and T. Sato, {\it The D
  coefficient in neutron beta decay in effective field theory},
  Phys. Lett. B {\bf 677}, 109 (2009); \\ DOI:
  https://doi.org/10.1016/j.physletb.2009.04.088.

\bibitem{Sirlin1967} A. Sirlin, {\it General properties of the
  electromagnetic corrections to the beta decay of a physical
  nucleon}, Phys. Rev. {\bf 164}, 1767
  (1967); \\ DOI:https://doi.org/10.1103/PhysRev.164.1767.

\bibitem{Shann1971} R. T. Shann, {\it Electromagnetic effects in the
  decay of polarized neutrons}, Nuovo Cimento A {\bf 5}, 591
  (1971).\\ DOI: https://doi.org/10.1007/BF02734566.

\bibitem{Sirlin1978} A. Sirlin, {\it Current algebra formulation of
  radiative corrections in gauge theories and the universality of the
  weak interactions}, Rev. Mod. Phys. {\bf 50}, 573 (1978); \\ DOI:
  https://doi.org/10.1103/RevModPhys.50.573.


\bibitem{Sirlin1986} W. J. Marciano and A. Sirlin, {\it Radiative
  corrections to $\beta$ decay and the possibility of a fourth
  generation}, Phys. Rev. Lett.  {\bf 56}, 22 (1986); \\ DOI:
  https://doi.org/10.1103/PhysRevLett.56.22.

\bibitem{Sirlin2004} A. Czarnecki, W. J. Marciano, and A. Sirlin, {\it
  Precision measurements and CKM unitarity}, Phys. Rev. D {\bf 70},
  093006 (2004);\\ DOI: 10.1103/PhysRevD.70.093006.

\bibitem{Sirlin2006} W. J. Marciano and A. Sirlin, {\it Improved
  calculation of electroweak radiative corrections and the value of
  $V(ud)$}, Phys. Rev. Lett.  {\bf 96}, 032002 (2006); \\ DOI:
  10.1103/PhysRevLett.96.032002.

\bibitem{Seng2018} Ch.-Y. Seng, M. Gorchtein, H. H. Patel, and
  M. J. Ramsey-Musolf, {\it Reduced hadronic uncertainty in the
    determination of $V_{ ud}$}, Phys. Rev. Lett. {\bf 121}, 241804
  (2018); \\ DOI: 10.1103/PhysRevLett.121.241804; arXiv:1807.10197
  [hep-ph].

\bibitem{Seng2018a} Ch.-Y. Seng, M. Gorchtein, and
  M. J. Ramsey-Musolf, {\it Dispersive evaluation of the inner
    radiative correction in neutron and nuclear beta decay},
  Phys. Rev. D {\bf 100}, 013001(2019); \\ DOI:
  10.1103/PhysRevD.100.013001; arXiv:1812.03352 [nucl-th].
  
\bibitem{Sirlin2019} A. Czarnecki, W. J. Marciano, and A. Sirlin, {\it
  Radiative corrections to neutron and nuclear beta decays revisited},
  Phys. Rev. D {\bf 100}, 073008 (2019);\\ DOI:
  10.1103/PhysRevD.100.073008.

\bibitem{Hayen2020} L. Hayen, {\it Standard Model $(\alpha)$
  renormalization of $g_A$ and its impact on new physics searches},
  arXiv: 2010.07262 [hep-ph]. 

\bibitem{Ebel1957} M. E. Ebel and G. Feldman, {\it Further remarks on
  Coulomb corrections in allowed beta transitions}, Nucl. Phys. {\bf
  4}, 213 (1957); \\ DOI:
  https://doi.org/10.1016/0029-5582(87)90020-4.
  

\bibitem{Feynman1958} R. P. Feynman and M. Gell--Mann, {\it Theory of
  Fermi interaction}, Phys. Rev.{\bf 109}, 193 (1958); \\ DOI:
  https://doi.org/10.1103/PhysRev.109.193.

\bibitem{Lee1956a} T. D. Lee and C. N. Yang, {\it
    Charge conjugation, a new quantum number $G$ , and selection rules
    concerning a nucleon anti-nucleon system}, Nuovo Cimento {\bf 10},
  749 (1956); \\ DOI: https://doi.org/10.1007/BF02744530.

\bibitem{Weinberg1958} S. Weinberg, {\it Charge symmetry of weak
  interactions}, Phys. Rev. {\bf 112}, 1375 (1958); \\ DOI:
  https://doi.org/10.1103/PhysRev.112.1375.


\bibitem{Wilkinson1970} D. H. Wilkinson and B. E. F. Macfield, {\it
  The numerical evaluation of radiative corrections of order $\alpha$
  to allowed nuclear $\beta$--decay}, Nucl. Phys. A {\bf 158}, 110
  (1970); \\ DOI: https://doi.org/10.1016/0375-9474(70)90055-2.


\bibitem{Berman1958} S. M. Berman, {\it Radiative corrections to muon
  and neutron decay}, Phys. Rev. {\bf 112}, 267 (1958); \\ DOI:
  https://doi.org/10.1103/PhysRev.112.267.

\bibitem{Kinoshita1959} T. Kinoshita and A. Sirlin, {\it Radiative
  corrections to Fermi interactions}, Phys. Rev. {\bf 113}, 1652
  (1959); \\ DOI: https://doi.org/10.1103/PhysRev.113.1652.


\bibitem{Ivanov2017b} A. N. Ivanov, R. H\"ollwieser, N. I. Troitskaya,
  M. Wellenzohn, and Ya. A. Berdnikov, {\it Precision theoretical
    analysis of neutron radiative beta decay to order
    $O(\alpha^2/\pi^2)$}, Phys. Rev. D {\bf 95}, 113006 (2017);
  \\ DOI: https://doi.org/10.1103/PhysRevD.95.113006.


\bibitem{Herczeg2001} P. Herczeg, {\it Beta decay beyond the standard
  model}, Progr. Part. Nucl. Phys. {\bf 46}, 413 (2001); \\ DOI:
  https://doi.org/10.1016/S0146-6410(01)00149-1.

  
\bibitem{Severijns2006} N. Severijns, M. Beck, and O. Naviliat-Cuncic,
  {\it Tests of the standard electroweak model in nuclear beta decay},
  Rev. Mod. Phys. {\bf 78}, 991 (2006); \\ DOI:
  https://doi.org/10.1103/RevModPhys.78.991.


\bibitem{Gardner2001} S. Gardner and C. Zhang, {\it Sharpening
  low-energy, Standard-Model tests via correlation coefficients in
  neutron beta decay}, Phys. Rev. Lett. {\bf 86}, 5666 (2001); \\ DOI:
  https://doi.org/10.1103/PhysRevLett.86.5666.

\bibitem{Gardner2013} S. Gardner and B. Plaster, {\it Framework for
  maximum likelihood analysis of neutron beta decay observables to
  resolve the limits of the V - A law}, Phys. Rev. C {\bf 87}, 065504
  (2013); The contribution to 4th International Conference on Particle
  Physics and Astrophysics (ICPPA-2018) 22–26 October 2018, Moscow,
  Russian Federation;\\ DOI: https://doi.org/10.1103/PhysRevC.87.06550
   

\bibitem{Kozela2009} A. Kozela, G. Ban, A. Bia\l ek, K. Bodek,
  P. Gorel, K. Kirch, St. Kistryn, M. Ku\'zniak, O. Naviliat-Cuncic,
  J. Pulut, N. Severijns, E. Stephan, and J. Zejma (the nTRV
  Collaboration), {\it Measurement of the transverse polarization of
    electrons emitted in free-neutron decay}, Phys. Rev. Lett. {\bf
    102}, 172301 (2009); \\ DOI:
  https://doi.org/10.1103/PhysRevLett.102.172301.
  
\bibitem{Kozela2012} A. Kozela, G. Ban, A. Bia\l ek, K. Bodek,
  P. Gorel, K. Kirch, St. Kistryn, O. Naviliat-Cuncic, N. Severijns,
  E. Stephan, and J. Zejma (the nTRV Collaboration), {\it Measurement
    of the transverse polarization of electrons emitted in free
    neutron decay}, Phys. Rev. C {\bf 85}, 045501 (2012); \\ DOI:
  https://doi.org/10.1103/PhysRevC.85.045501.

\bibitem{Wichers1987} V. A. Wichers, T. R. Hageman, J. van Klinken,
  and H. W. Wilschut, {\it Bounds on right-handed currents from
    nuclear beta decay}, Phys. Rev. Lett. {\bf 58}, 1821 (1987);
  \\ DOI: https://doi.org/10.1103/PhysRevLett.58.1821.

\bibitem{Carnoy1990} A. S. Carnoy, J. Deutsch, T. A. Girard, and
  R. Prieels, {\it Limits on nonstandard weak currents from the
    positron decays of ${^{14}}{\rm O}$ and ${^{10}}{\rm C}$},
  Phys. Rev. Lett. {\bf 65}, 3249 (1990); \\ DOI:
  https://doi.org/10.1103/PhysRevLett.65.3249.

\bibitem{Carnoy1991} A. S. Carnoy, J. Deutsch, T. A. Girard, and
  R. Prieels, {\it Limits on nonstandard weak currents from the
    polarization of ${^{14}}{\rm O}$ and ${^{10}}{\rm C}$ decay
    positrons}, Phys. Rev. C {\bf 43}, 2825 (1991); \\ DOI:
  https://doi.org/10.1103/PhysRevC.43.2825.

\bibitem{Gluck1995} F. Gl\"uck, I. Jo\'o, and J. Last, {\it Measurable
  parameters of neutron decay}, Nucl. Phys. A {\bf 593}, 125 (1995);
  \\ DOI: https://doi.org/10.1016/0375-9474(95)00354-4.
  
\bibitem{Serebrov1995} I. A. Kuznetzov, A. P. Serebrov,
  I. V. Stepanenko, A. V. Alduschenkov, M. S. Lasakov, A. A. Kokin,
  Yu. A. Mostovoi, B. G. Yerozolimsky, and M. S. Dewey, {\it
    Measurements of the antineutrinos Spin asymmetry in beta decay of
    the neutron and restrictions on the mass of a right-handed gauge
    boson}, Phys. Rev. Lett. {\bf 75}, 794 (1995); \\ DOI:
  https://doi.org/10.1103/PhysRevLett.75.794.

\bibitem{Serebrov1998} A. P. Serebrov, I. A. Kuznetsov,
  I. V. Stepanenko, A. V. Aldushchenkov, M. S. Lasakov,
  Yu. A. Mostovoi, B. G. Erozolimskii, M. S. Dewey, F. E. Wietfeldt,
  O. Zimmer, and H. B\"orner, {\it Measurement of the anti-neutrino
    escape asymmetry with respect to the spin of the decaying
    neutron}, Zh. Exsp. Teor. Fiz. {\bf 113}, 1963 (1998),
  J. Exp. Theor.Phys. {\bf 86}, 1074 (1998); \\ DOI:
  https://doi.org/10.1134/1.558574.

\bibitem{Schumann2007} M. Schumann, T. Soldner, M. Deissenroth,
  F. Gl\"uck, J. Krempel, M. Kreuz, B. M\"arkisch, D. Mund,
  A. Petoukhov, and H. Abele, {\it Measurement of the neutrino
    asymmetry parameter B in neutron decay}, Phys. Rev. Lett. {\bf
    99}, 191803 (2007); \\ DOI:
  https://doi.org/10.1103/PhysRevLett.99.191803.

\bibitem{Mitchell1949} K. Mitchell, {\it XXXII. Tables of the
  functions $- \int^z_0\frac{{\rm log}(1 - y)}{y}\,dy$, with an
  account for some properties of this and related functions}, The
  London, Edinburgh, and Dublin Philosophical Magazine and Journal of
  Science, {\bf 40}, 351 (1949); \\ DOI:
  https://doi.org/10.1080/14786444908561256.
  

\end{thebibliography}
\end{document}